\documentclass[a4paper,USenglish]{lipics-v2019}

\PassOptionsToPackage{hyphens}{url}\usepackage{hyperref}
\usepackage[nonatbib,esvect,lipics]{bec} %

\newcommand{\refappendix}[1]{the extended version~\cite{verivitatr}}

\usepackage{subcaption}

\setboolean{showjedi}{false}
\setboolean{showcomments}{false}
\usepackage{tikz}
\usepackage[underline=false]{pgf-umlsd}
\usepackage{sidecap}
\usetikzlibrary{arrows,automata,shapes,calc,decorations.pathreplacing,calc,trees,positioning,arrows,chains,shapes.geometric,%
	decorations.pathreplacing,decorations.pathmorphing,shapes,%
	matrix,shapes.symbols}

\usepackage{todonotes}
\usepackage{thmtools,thm-restate}
\usepackage{multirow}

\DeclareCaptionSubType*{figure}

\title{Lifestate: Event-Driven Protocols and Callback Control Flow}

\author{Shawn Meier}{University of Colorado Boulder, USA}{}{}{}
\author{Sergio Mover}{\'{E}cole Polytechnique, France}{}{}{}
\author{Bor-Yuh Evan Chang}{University of Colorado Boulder, USA}{}{}{}

\authorrunning{Shawn Meier, Sergio Mover, and Bor-Yuh Evan Chang}

\Copyright{Shawn Meier, Sergio Mover, and Bor-Yuh Evan Chang}



\EventEditors{Alastair F. Donaldson}
\EventNoEds{1}
\EventLongTitle{33rd European Conference on Object-Oriented Programming (ECOOP 2019)}
\EventShortTitle{ECOOP 2019}
\EventAcronym{ECOOP}
\EventYear{2019}
\EventDate{July 15--19, 2019}
\EventLocation{London, United Kingdom}
\EventLogo{}
\SeriesVolume{134}
\ArticleNo{4}

\begin{document}
\nolinenumbers
\maketitle

\begin{abstract}
%
%
%
Developing interactive applications (apps) against event-driven software frameworks such as Android is notoriously difficult.
To create apps that behave as expected, developers must follow complex and often implicit \emph{asynchronous programming protocols}.
Such protocols intertwine the proper registering of callbacks to receive control from the framework with appropriate application-programming interface (API) calls that in turn affect the set of possible future callbacks.
An app violates
the protocol
when, for example, it calls a particular API method in a state of the framework where such a call is invalid.
%
%
What makes
automated reasoning hard in this domain is
largely what makes programming apps against such frameworks hard: the
specification of the protocol is unclear, and the control flow is complex, asynchronous, and higher-order.
%
%
In this paper, we tackle the problem of specifying and modeling
event-driven application-programming protocols.
In particular, we formalize a core meta-model that captures the
dialogue between event-driven frameworks and application callbacks.
Based on this
meta-model, we define a language called \emph{lifestate} that permits precise and formal descriptions of 
application-programming protocols and the callback control flow
imposed by the event-driven framework.
Lifestate unifies modeling what app callbacks can expect of the framework with specifying rules the app must respect when calling into the framework. 
In this way, we effectively combine lifecycle constraints and typestate rules.
To evaluate the effectiveness of lifestate modeling,
we provide a dynamic verification algorithm that takes as input a trace of execution of an app and a lifestate protocol specification to either produce a trace witnessing a protocol violation or a proof that no such trace is realizable.

\end{abstract}

\input{macros}

\newsavebox{\SBoxOnClick}
\codejmk[\small]{\SBoxOnClick}{(l:OnClickListener).onClick(b:Button)}
\newsavebox{\SBoxExecute}
\codejmk[\small]{\SBoxExecute}{(t:AsyncTask).execute()}
\newsavebox{\SBoxSetOnClickListener}
\codejmk[\small]{\SBoxSetOnClickListener}{(b:Button).setOnClickListener(l:OnClickListener)}
\newsavebox{\SBoxSetContentView}
\codejmk[\small]{\SBoxSetContentView}{(a:Activity).setContentView(v:View)}
\newsavebox{\SBoxOnCreate}
\codejmk[\small]{\SBoxOnCreate}{(a:Activity).onCreate()}
\newsavebox{\SBoxOnStart}
\codejmk[\small]{\SBoxOnStart}{(a:Activity).onStart()}
\newsavebox{\SBoxOnResume}
\codejmk[\small]{\SBoxOnResume}{(a:Activity).onResume()}
\newsavebox{\SBoxOnPause}
\codejmk[\small]{\SBoxOnPause}{(a:Activity).onPause()}
\newsavebox{\SBoxSetEnabledTrueCall}
\codejmk[\small]{\SBoxSetEnabledTrueCall}{(b:Button).setEnabled(true)}
\newsavebox{\SBoxSetEnabledFalseCall}
\codejmk[\small]{\SBoxSetEnabledFalseCall}{(b:Button).setEnabled(false)}

\section{Introduction}
\label{sec:introduction}

\JEDI{App must respects a framework-dictated application programming protocol.}%
We consider the essential problem of checking that an application (app) programmed against an event-driven framework respects the required application-programming protocol.
In such frameworks, apps implement \emph{callback} interfaces
so that the app is
notified when an \emph{event} managed by the framework occurs (e.g., a user-interface~(UI) button is pressed).
The app may then delegate back to the framework through calls to the application programming interface~(API),
which we term \emph{callin} by analogy to callback.
To develop working apps, the programmer must reason about hidden
\emph{callback control flow} and often implicit asynchronous programming protocols.

%
%
%
\newsavebox{\SBoxDismiss}
\begin{lrbox}{\SBoxDismiss}\small
\begin{lstlisting}[language=Java]
try { progress.dismiss(); } catch ($\mbox{\relsize{-1}IllegalArgumentException}$ ignored) {} // race condition?
\end{lstlisting}
\end{lrbox}
%
\begin{figure}\centering
\usebox{\SBoxDismiss}%
\caption{A protocol ``fix''~\cite{boyle-dismissProgress}. The \codej{dismiss} call throws an exception if called in an invalid state.}%
\label{fig:dismiss}%
\end{figure}

%
Couple difficult reasoning about the space of possible control flow between
callbacks with insufficient framework documentation, and
it is unsurprising to find some questionable ``fixes'' for protocol violations.
In \figref{dismiss}, we show a snippet found on GitHub.
The ``race condition?'' comment is quoted directly from the app
developer.
%
%
%
%
The same asynchronous, implicitly defined, control flow that make it difficult for the app developer to reason about his app is also what makes verifying the absence of such protocol violations hard.
%


\tikzset{
  state/.style = {
    rectangle,
    rounded corners,
    draw=black, very thick,
    minimum height=1.5em,
    inner sep=2pt,
    text centered,
    },
      lifearrow/.style={->, >=latex', line width = 1},
      missingarrow/.style={->,>=latex',line width = 1,densely dashed,red},
      plusplusarrow/.style={->,>=latex',line width = 1.2,densely dotted,blue},
      missingline/.style={-,>=latex',line width = 1,denesly dashed,red}
 }
\newcommand{\lifearrow}{\tikz{\draw[lifearrow]($(0cm,0.5cm)$) node [right] (a){} -- ($(0.5cm,0.5cm)$);}}
\newcommand{\plusplusarrow}{\tikz{\draw[plusplusarrow]($(0cm,0.25cm)$) node [right] (a){} -- ($(0.5cm,0.25cm)$);}}
\newcommand{\missingarrow}{\tikz{\draw[missingarrow]($(0cm,0.25cm)$) node [right](a){} -- ($(0.5cm,0.25cm)$);}}

\begin{figure}[b]\centering
\includegraphics[scale=0.7]{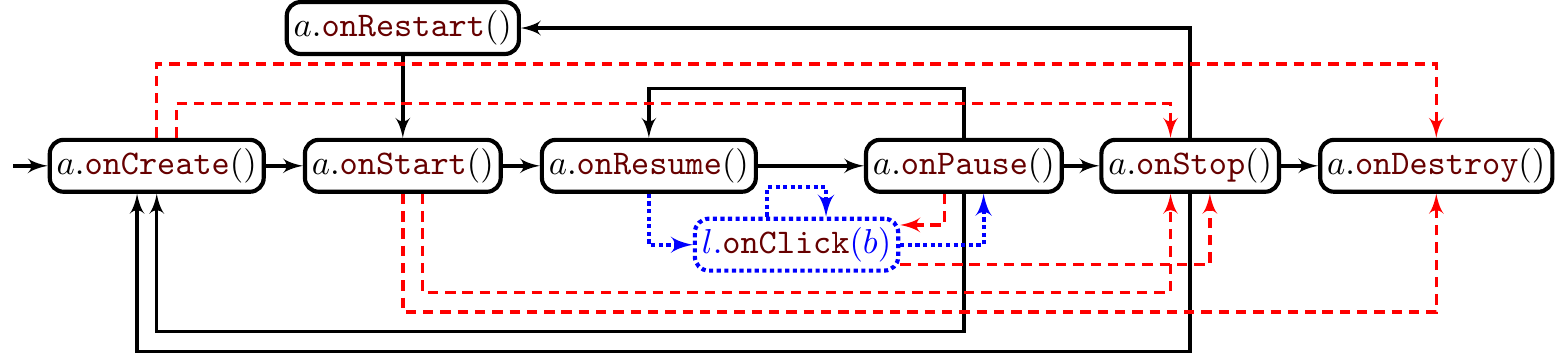}
\caption{The \codej{Activity} lifecycle automaton from the Android documentation~\cite{android-activity-lifecycle} (shown with solid, black edges \lifearrow). To capture callback control flow between ``related'' objects, such component lifecycles are often instantiated and refined with additional callbacks from other objects, such as a \codej{onClick} callback from the \codej{OnClickListener} interface (shown with dotted, blue edges \plusplusarrow).
But there are also less common callback control-flow paths that are often undocumented or easily missed, such as
the additional edges induced by an invocation in the app code of the \codej{finish} callin (shown as dashed, red edges \missingarrow).}
\label{fig:activitylifecycle}
\end{figure}

\JEDI{The existing ways of modeling the callback control flow are not adequate to reason about protocol violations}

In this paper, we focus on the problems of specifying event-driven protocols (i.e., specifying when the invocation of a callin in the app code causes a protocol violation) and modeling the callback control flow (i.e., modeling the possible executions of callbacks).

\subparagraph{Lifecycle Automata are Insufficient for Modeling Callback Control Flow.}
%
%
\emph{Lifecycle} automata are a common representation used to model
callback control flow
that is both central to Android
documentation~\cite{android-activity-lifecycle,xxv-androidlifecycle} and prior Android analysis techniques---both static and dynamic ones  (e.g., \cite{DBLP:conf/pldi/ArztRFBBKTOM14,DBLP:conf/pldi/MaiyaKM14,DBLP:conf/oopsla/BlackshearCS15}).
In \figref{activitylifecycle}, we show a lifecycle automaton for the \codej{Activity} class of the Android framework. The black, solid edges are the edges present in the Android documentation~\cite{android-activity-lifecycle} showing common callback control flow.
These edges capture, for example, that the app first receives the \codej{onStart} callback before entering a cycle between the \codej{onResume} and the \codej{onPause} callbacks. But this clean and simple class-based model quickly becomes insufficient when we look deeper.

First, there are complex relationships between the callbacks on ``related'' objects. For example, an \codej{OnClickListener} object $l$ with an \codej{onClick} callback may be ``registered'' on a \codej{View} object $v$ that is ``attached'' to an \codej{Activity} object $a$. Because of these relationships, the callback control flow we need to capture is somewhat described by modifying the lifecycle automaton for \codej{Activity} $a$ with the additional blue, dotted edges to and from \codej{onClick} (implicitly for \codej{OnClickListener} $l$) in \figref{activitylifecycle}. This modified lifecycle encodes framework-specific knowledge that the \codej{OnClickListener} $l$'s \codej{onClick} callback happens only in the ``active'' state of \codej{Activity} $a$ between its \codej{onResume} and \codej{onPause} callbacks, which typically requires a combination of static analysis on the app and hard-coded rules to connect callbacks on additional objects such as \codej{OnClickListener}s to component lifecycles such as \codej{Activity}. We refer to such callback control-flow models based on such refined lifecycle automatons as \lifecyclepp{} models.


\JEDI{Lifecycle is not enough for protocol verification}
Second, there are less common framework-state changes that are difficult to capture soundly and precisely. For example, an analysis that relies on a callback control-flow model that does not consider the intertwined effect of a \codej{finish} call may be unsound.
%
The red, dashed edges represent callback control flow that are not documented  (and thus missing from typical callback control flow models).
Each one of these edges specifies different possible callback control flow that the framework imposes depending on
\emph{if and when} the app invokes the \codej{finish} callin inside one of the \codej{Activity}'s callbacks.
%
Of course, the lifecycle automaton can be extended to include these red edges. However, this lifecycle automaton is now quite imprecise in the common case because it does not express precisely when certain callback control-flow paths are spurious (i.e., depending on where \codej{finish} is not called).
%
%
%
%
\Figref{activitylifecycle} illustrates why developing callback control flow models is error prone: the effect of calls to \codej{finish} are subtle and poorly understood.
%
%

It is simply too easy to miss possible callback control flow---an observation also made by
Wang et al.~\cite{DBLP:conf/pldi/WangZR16} about lifecycle models.
While lifecycle automata are useful for conveying the intuition of callback control flow, they are often insufficiently precise and easily unsound.

In this paper, we re-examine the process of modeling callback control flow.
In prior work, modeling callback control flow was almost always a secondary concern in service to, and often built into, a specific program analysis where the analysis abstraction may reasonably mask unsound callback control flow.
Instead, we consider modeling callback control flow independent of any analysis abstraction---we identify and formalize the key aspects to effectively model event-driven application-programming protocols at the app-framework interface, such as the effect of callin and callback invocations on the subsequent callback control flow,
This first-principles approach enables us to \emph{validate} callback control-flow soundness with real execution traces against the event-driven framework implementation.
It is through this validation step that we discovered the red, dashed edges in \figref{activitylifecycle}.

%
%

\JEDI{A similar problem happens for specifying the protocol violations (typestates)}

\subparagraph{Contributions.}
We make the following contributions:
\begin{itemize}\itemsep 0pt
  \item We identify essential aspects of event-driven control flow and
    application-programming protocols to formalize a core abstract machine model $\semname$ (\secref{concrete}). This model provides a formal basis for thinking about event-driven frameworks and their application-programming protocols.
  \item We define a language for simultaneously capturing event-driven
    application-programming protocols and callback control flow called \emph{lifestates}, which both \emph{model} what callback invocations an app can expect from the framework and \emph{specify} rules the app must respect when calling into the framework (\secref{specification}).
Intuitively, lifestates offer the ability to specify traces of the event-driven program in terms of an abstraction of the observable interface between the framework and the app. And thus, this definition leads to a methodology for empirically validating lifestate models against actual interaction traces.
%
  \item We define lifestate validation and dynamic lifestate verification. And then, we encode them as model checking problems (\secref{dynamic}). Given an app-framework interaction trace and a lifestate model, validation checks that the trace is in the abstraction of the observable interface defined by the model. This validation can be done with corpora of traces recorded from any set of apps interacting with the same framework because, crucially, the lifestate model speaks only about the app-framework interface. Then, given a trace, dynamic lifestate verification attempts to prove the absence of a rearrangement of the recorded events that could cause a protocol violation. Rearranging the execution trace of events corresponds to exploring a different sequence of external inputs and hence discovering possible protocol violations not observed in the original trace.
%
%
  \item We implement our model validation and trace verification approach in a tool called \toolname{} and use it to empirically evaluate the soundness and precision of callback control flow models of Android (\secref{expeval}). Our results provide evidence for the hypotheses that lifecycle models, by themselves, are insufficiently precise to verify Android apps as conforming to the specified protocols, that model validation on large corpora of traces exposes surprising unsoundnesses, and that lifestates are indeed useful.
\end{itemize}

\section{Overview: Specifying and Modeling Lifestates}
\label{sec:overview}
Here, we illustrate the challenges in specifying and modeling event-driven application-programming protocols. In particular, we motivate the need for lifestates that permit specifying the intertwined effect of callin and callback invocations. We show that even if an app is buggy, it can be difficult to witness the violation of the Android application programming protocol.  Then, more importantly, we show how an appropriate fix is both subtle to reason about and requires modeling the complex callback control flow that depends on the previous execution of not only the callbacks but also the callins.
%
%
%

\newsavebox{\SBoxSetEnabledFalse}
\begin{lrbox}{\SBoxSetEnabledFalse}\small
\codej{button.setEnabled(false)}
\end{lrbox}
\newsavebox{\SBoxFeedRemoverExampleRemoverActivity}
\begin{lrbox}{\SBoxFeedRemoverExampleRemoverActivity}\small
\lststopn\begin{lstlisting}[language=Java,alsolanguage=exthighlighting,style=number,name=Bugg]
class RemoverActivity extends Activity {
 FeedRemover remover;
 void onCreate() {#\lstbeginn#
  Button button = $\ldots$;
  remover = new FeedRemover(this);#\label{line-feedremover-firstalloc}#
  button.setOnClickListener(
   new OnClickListener() {#\label{line-feedremover-listener}\lststopn#
    void onClick(View view) {#\lststartn#
     remover.execute();   #\danger\label{line-feedremover-danger}\lststopn#
    }
   });
 }
}
\end{lstlisting}
\end{lrbox}

\newsavebox{\SBoxFeedRemoverExampleAsyncTask}
\begin{lrbox}{\SBoxFeedRemoverExampleAsyncTask}\small
\begin{lstlisting}[language=Java,alsolanguage=exthighlighting,style=number,name=Bugg]
class FeedRemover extends AsyncTask {
 RemoverActivity activity;
 void doInBackground() {
  $\ldots \text{\sl remove feed}\;\ldots$
 }
 void |cb:onPostExecute|() {
  // return to previous activity#\lststartn#
  activity.finish(); #\label{line-feedremover-alloc}\lststopn#
 }
}#\lststartn#
\end{lstlisting}
\end{lrbox}

\newsavebox{\SBoxFeedRemoverFixedOne}
\begin{lrbox}{\SBoxFeedRemoverFixedOne}\small
\lststopn\begin{lstlisting}[language=Java,alsolanguage=exthighlighting,style=number]
class RemoverActivity extends Activity {
 FeedRemover remover;
 void onCreate() {#\lstbeginn#
  Button button = $\ldots$;
  remover = new FeedRemover(this);#\label{line-feedremover-firstalloc-fixed}#
  button.setOnClickListener( #\label{line-feedremover-register}#
   new OnClickListener() {#\label{line-feedremover-listener-fixed}\lststopn#
    void onClick(View view) {#\lststartn#
\end{lstlisting}
\end{lrbox}

\newsavebox{\SBoxFeedRemoverFixedTwo}
\begin{lrbox}{\SBoxFeedRemoverFixedTwo}\small
\begin{lstdiffplus}[language=Java,alsolanguage=exthighlighting,style=number,linewidth=0.4\linewidth]
+    #\usebox{\SBoxSetEnabledFalse}#; #\label{line-feedremover-disable-fixed}#
\end{lstdiffplus}
\end{lrbox}

\newsavebox{\SBoxFeedRemoverFixedThree}
\begin{lrbox}{\SBoxFeedRemoverFixedThree}\small
\begin{lstcont}[language=Java,alsolanguage=exthighlighting,style=number]
     remover.execute();   #\danger\label{line-feedremover-danger-fixed}\lststopn#
    }
   });
 }
}

class FeedRemover extends AsyncTask {
 RemoverActivity activity;
 void doInBackground() {
  $\ldots\;\text{\sl remove feed}\;\ldots$
 }
 void |cb:onPostExecute|() {
  // return to previous activity#\lststartn#
  activity.finish(); #\label{line-feedremover-alloc-fixed}\lststopn#
\end{lstcont}
\end{lrbox}
\newsavebox{\SBoxFeedRemoverFixedFour}
\newsavebox{\SBoxFeedRemoverFixedFive}
\begin{lrbox}{\SBoxFeedRemoverFixedFive}\small
\begin{lstcont}[language=Java,alsolanguage=exthighlighting,style=number]
 }
}#\lststartn#
\end{lstcont}
\end{lrbox}
\newsavebox{\SBoxAsyncTaskInterface}
\begin{lrbox}{\SBoxAsyncTaskInterface}\small
\begin{lstlisting}[language=Java,alsolanguage=exthighlighting]
abstract class AsyncTask {
void |cb:onPostExecute|() { }
abstract void doInBackground();
final void execute() { $\ldots$ }
}
\end{lstlisting}
\end{lrbox}
\newsavebox{\SBoxAsyncExecute}\codejmk[\small]{\SBoxAsyncExecute}{(t:AsyncTask).execute()}
\newsavebox{\SBoxAsyncExecuteNoType}\codejmk[\small]{\SBoxAsyncExecuteNoType}{t.execute()}
\newsavebox{\SBoxAsyncPostExecute}\codejmk[\small]{\SBoxAsyncPostExecute}{t.|cb:onPostExecute|()}
\newsavebox{\SBoxButtonEnable}\codejmk[\small]{\SBoxButtonEnable}{(b:Button).setEnabled(true)}
\newsavebox{\SBoxButtonDisable}\codejmk[\small]{\SBoxButtonDisable}{(b:Button).setEnabled(false)}
\newsavebox{\SBoxListenerOnClick}\codejmk[\small]{\SBoxListenerOnClick}{(l:OnClickListener).onClick(b)}
\newsavebox{\SBoxAsyncInit}\codejmk[\small]{\SBoxAsyncInit}{(t:AsyncTask).<init>()}
\newsavebox{\SBoxFeedRemoverTrace}
\begin{lrbox}{\SBoxFeedRemoverTrace}\footnotesize
\begin{sequencediagram}
\newthread{f}{Framework\vphantom{App}}
\newinst[5.5]{a}{App\vphantom{Framework}}
\begin{scope}[>=triangle 60]
\begin{sdblock}{\fmtevt{Create}}{}
  \begin{messcall}{f}{ \codej{(a:Activity).onCreate()} }{a}{}
    \begin{messcall}{a}{ \codej{(t:AsyncTask).|ci:<init>|()} }{f}{}
      \node[right=1mm] at (cf\thecallevel) {\lstnumberstyle\ref{line-feedremover-firstalloc}};
    \end{messcall}\prelevel
    \begin{messcall}{a}{
        \codej{(b:Button).setOnClickListener(l:OnClickListener)}
      }{f}{}
      \node[right=1mm] at (cf\thecallevel) {\lstnumberstyle\ref{line-feedremover-listener}};
    \end{messcall}\prelevel
  \end{messcall}\prelevel
\end{sdblock}
\path (blocktitle.west) + (2,0) node {\makebox[0pt][l]{Create \codej{Activity} \codej{a}.}};
\begin{sdblock}{\fmtevt{Click}}{}
  \begin{call}{f}{ \codej{(l:OnClickListener).onClick(b:Button)} }
    {a}{ \codej{(l:OnClickListener).onClick(b:Button)} }
    \begin{call}{a}{ \codej{(t:AsyncTask).execute()} }
      {f}{ \codej{(t:AsyncTask) = (t:AsyncTask).execute()} }
      \node[right=1mm] at (cf\thecallevel) {\lstnumberstyle\ref{line-feedremover-danger}};
    \end{call}
  \end{call}
\end{sdblock}
\path (blocktitle.west) + (2,0) node {\makebox[0pt][l]{User clicks on \codej{Button} \codej{b}.}};
\makeatletter
\begin{pgfonlayer}{umlsd@background}
  \filldraw[fill=yellow!5] (se) rectangle (nw);
\end{pgfonlayer}
\makeatother
\begin{sdblock}{\fmtevt{PostExecute}}{}
  \begin{messcall}{f}{ \codej{(t:AsyncTask).|cb:onPostExecute|()} }{a}{}
    \begin{messcall}{a}{ \codej{(a:Activity).finish()} }{f}{}
      \node[right=1mm] at (cf\thecallevel) {\lstnumberstyle\ref{line-feedremover-alloc}};
    \end{messcall}\prelevel
  \end{messcall}\prelevel
\end{sdblock}
\path (blocktitle.west) + (2,0) node {\makebox[0pt][l]{Finish \codej{AsyncTask} \codej{t}.}};
\end{scope}
\node (cb) at (2.2,0.3) {};
\node (cbend) [right of=cb,label=right:calling a callback] {};
\node (cbretend) at (2.2,0) {};
\node (cbret) [right of=cbretend,label=right:returning from a callback] {};
\node (ciend) at (2.2,-0.5) {};
\node (ci) [right of=ciend,label=right:calling a callin] {};
\node (ciret) at (2.2,-0.8) {};
\node (ciretend) [right of=ciret,label=right:returning from a callin] {};
\draw[->,>=triangle 60] (cb) -- (cbend);
\draw[dashed,->,>=angle 60] (cbret) -- (cbretend);
\draw[->,>=triangle 60] (ci) -- (ciend);
\draw[dashed,->,>=angle 60] (ciret) -- (ciretend);
\end{sequencediagram}
\end{lrbox} 
\newsavebox{\SBoxFeedRemoverTraceBuggy}
\begin{lrbox}{\SBoxFeedRemoverTraceBuggy}\footnotesize
\begin{sequencediagram}
\newthread{f}{Framework\vphantom{App}}
\newinst[5.5]{a}{App\vphantom{Framework}}
\begin{scope}[>=triangle 60]
\node (spacer) at (7,0.2) {};
\node (spacerText) [right of=spacer,label=] {};
\begin{sdblock}{\fmtevt{Create}}{}
  \begin{messcall}{f}{ \codej{(a:Activity).onCreate()} }{a}{}
    \begin{messcall}{a}{ \codej{(t:AsyncTask).|ci:<init>|()} }{f}{}
      \node[right=1mm] at (cf\thecallevel) {\lstnumberstyle\ref{line-feedremover-firstalloc}};
    \end{messcall}\prelevel
    \begin{messcall}{a}{ \codej{(b:Button).setOnClickListener(l:OnClickListener)} }{f}{}
      \node[right=1mm] at (cf\thecallevel) {\lstnumberstyle\ref{line-feedremover-listener}};
    \end{messcall}\prelevel
  \end{messcall}\prelevel
\end{sdblock}
\begin{sdblock}{\fmtevt{Click}}{}
  \begin{messcall}{f}{ \codej{(l:OnClickListener).onClick(b:Button)} }{a}{}
    \begin{messcall}{a}{ \codej{(t:AsyncTask).execute()} }{f}{}
      \node[right=1mm] at (cf\thecallevel) {\lstnumberstyle\ref{line-feedremover-danger}};
    \end{messcall}\prelevel
  \end{messcall}\prelevel
\end{sdblock}
\begin{sdblock}{\fmtevt{Click}}{}
  \begin{messcall}{f}{ \codej{(l:OnClickListener).onClick(b:Button)} }{a}{}
    \begin{messcall}{a}{ \codej{(t:AsyncTask).execute()} }{f}{}
      \node[right=1mm] at (cf\thecallevel) {\lstnumberstyle\ref{line-feedremover-danger}};
      \newcommand{\Cross}{\tikz [x=3.0ex,y=3.0ex,line
        width=1ex, red] \draw (0,0) -- (1,1) (0,1) -- (1,0);}
      \node[] (crosssign) at (ct\thecallevel) {\makebox[0pt]{\Cross}};
    \end{messcall}\prelevel
  \end{messcall}\prelevel
\end{sdblock}
\begin{sdblockx}{}{ \parbox{5.7cm}{The \codej{AsyncTask} \codej{t} is still running, so the \fmtevt{PostExecute} event has not yet happened.} }{dotted}\prelevel\prelevel
\messx{f}{}{a}{white}
\node[right=1mm] at (mess to) {\phantom{5}};
\end{sdblockx}
\end{scope}
\end{sequencediagram}
\end{lrbox} 
%
\newsavebox{\SBoxFeedRemoverTraceFixed}
\begin{lrbox}{\SBoxFeedRemoverTraceFixed}\footnotesize
\begin{sequencediagram}
\newthread{f}{Framework\vphantom{App}}
\newinst[5.5]{a}{App\vphantom{Framework}}
\begin{scope}[>=triangle 60]
\begin{sdblock}{\fmtevt{Create}}{}
  \begin{messcall}{f}{ \codej{(a:Activity).onCreate()} }{a}{}
    \path (cf\thecallevel) + (0,1.65) node[left=0.5em] {
      \begin{tabular}{@{}r@{}}
      \cbstyle\bf enabled callbacks\,$\circcheck$
      \\
      \cistyle\bf disallowed callins\,\danger
      \end{tabular}
    };
    \node[left=0.5em] at (cf\thecallevel) {\codej{a.onCreate()}\,$\circcheck$};
    \begin{messcall}{a}{ \codej{(t:AsyncTask).|ci:<init>|()} }{f}{}
      \node[left=0.5em] at (ct\thecallevel) {\phantom{\codej{t1.onPostExecute()}\,$\circcheck$}};
    \node[right=1mm] at (cf\thecallevel) {\lstnumberstyle\ref{line-feedremover-firstalloc-fixed}};
    \end{messcall}\prelevel
    \begin{messcall}{a}{ \codej{(b:Button).setOnClickListener(l:OnClickListener)} }{f}{}
      \node[left=0.5em] at (ct\thecallevel) {\codej{l.onClick(b)}\,$\circcheck$};
      \node[right=1mm] at (cf\thecallevel) {\lstnumberstyle\ref{line-feedremover-listener-fixed}};
    \end{messcall}\prelevel
  \end{messcall}\prelevel
\end{sdblock}
\begin{sdblock}{\fmtevt{Click}}{}
  \begin{messcall}{f}{ \codej{(l:OnClickListener).onClick(b:Button)} }{a}{}
    \node[left=0.5em] at (cf\thecallevel) {\codej{l.onClick(b)}\,$\circcheck$};
    \begin{messcall}{a}{ \codej{(b:Button).setEnabled(false)} }{f}{}
      \node[right=1mm] at (cf\thecallevel) {\lstnumberstyle\ref{line-feedremover-disable-fixed}};
    \end{messcall}\prelevel
    \begin{messcall}{a}{ \codej{(t:AsyncTask).execute()} }{f}{}
      \node[left=0.5em] at (ct\thecallevel) {
      \begin{tabular}{@{}r@{}}
          \codej{t.onPostExecute()}\,$\circcheck$
          \\
          \codej{t.execute()}\,\danger
          \end{tabular}
      };
      \node[right=1mm] at (cf\thecallevel) {\lstnumberstyle\ref{line-feedremover-danger-fixed}};
    \end{messcall}\prelevel
  \end{messcall}\prelevel
\end{sdblock}
\begin{sdblockx}{}{ The \fmtevt{Click} is not enabled, so it cannot happen here. }{dotted}\prelevel\prelevel
\messx{f}{}{a}{white}
\node[right=1mm] at (mess to) {\phantom{5}};
\end{sdblockx}
\begin{sdblock}{\fmtevt{PostExecute}}{}
  \begin{messcall}{f}{ \codej{(t:AsyncTask).|cb:onPostExecute|()} }{a}{}
    \node[left=0.5em] at (cf\thecallevel) {
      \begin{tabular}{@{}r@{}}
        \codej{t.onPostExecute()}\,$\circcheck$
        \\
        \codej{t.execute()}\,\danger
    \end{tabular}
    };
    \begin{messcall}{a}{ \codej{(a:Activity).finish()} }{f}{} 
    \node[right=1mm] at (cf\thecallevel) {\lstnumberstyle\ref{line-feedremover-alloc-fixed}};
    \end{messcall}\prelevel
  \end{messcall}\prelevel
\end{sdblock}
\end{scope}
\end{sequencediagram}
\end{lrbox} 

\newsavebox{\SBoxHighlighted}
\sbox{\SBoxHighlighted}{\tikz[baseline=(X.base)]{\node[fill=yellow!5,inner sep=0pt] (X) {\small highlighted};}}
\newsavebox{\SBoxCallbackArrow}
\sbox{\SBoxCallbackArrow}{\tikz{\draw[->,>=triangle 60] (0,0) -- (0.5,0);}}
\newsavebox{\SBoxCallinArrow}
\sbox{\SBoxCallinArrow}{\tikz{\draw[->,>=triangle 60] (0.5,0) -- (0,0);}}
\newsavebox{\SBoxCallinReturnArrow}
\sbox{\SBoxCallinReturnArrow}{\tikz{\draw[dashed,->,>=angle 60] (0,0) -- (0.4,0);}}
\newsavebox{\SBoxCallbackReturnArrow}
\sbox{\SBoxCallbackReturnArrow}{\tikz{\draw[dashed,->,>=angle 60] (0.4,0) -- (0,0);}}
\newsavebox{\SBoxActivationShading}
\sbox{\SBoxActivationShading}{\tikz{\draw[fill=gray!30] (0,0) rectangle +(0.75em,1ex);}}

\newcommand{\TheFeedRemoverTrace}{\scalebox{0.7}{\usebox{\SBoxFeedRemoverTrace}}}
\newcommand{\TheFeedRemoverTraceBuggy}{\scalebox{0.7}{\usebox{\SBoxFeedRemoverTraceBuggy}}}
\newcommand{\TheFeedRemoverExampleRemoverActivity}{\scalebox{1}{\usebox{\SBoxFeedRemoverExampleRemoverActivity}}}
\newcommand{\TheFeedRemoverExampleAsyncTask}{\scalebox{1}{\usebox{\SBoxFeedRemoverExampleAsyncTask}}}
\begin{figure}[b]
\begin{tabular*}{\linewidth}{@{\extracolsep{\fill}}ll@{}}
\begin{minipage}[b]{0.45\linewidth}
\TheFeedRemoverExampleRemoverActivity
\end{minipage}
&
\begin{minipage}[b][\heightof{\TheFeedRemoverExampleRemoverActivity}][t]{0.45\linewidth}
\TheFeedRemoverExampleAsyncTask
\end{minipage}
\end{tabular*}
\caption{An example app that violates the
protocol specified by the interaction of the Android framework
components \codej{AsyncTask}, \codej{Button}, and \codej{OnClickListener}.
On \reftxt{line}{line-feedremover-danger},
\codej{remover.execute()}
(marked with \danger) can throw an \codej{IllegalStateException} if the
\codej{remover} task is already running.%
}
\label{fig:feedremover-buggy-code}
\end{figure}

\newsavebox{\SBoxRemoverExecute}
\codejmk[\small]{\SBoxRemoverExecute}{remover.execute()}
\newsavebox{\SBoxOldTaskAlloc}
\codejmk[\small]{\SBoxOldTaskAlloc}{(t:AsyncTask).|ci:<init>|()}
\newsavebox{\SBoxNewTaskAlloc}
\codejmk[\small]{\SBoxNewTaskAlloc}{(t2:AsyncTask).|ci:<init>|()}
\newsavebox{\SBoxAFinish}
\codejmk[\small]{\SBoxAFinish}{activity.finish()}
\newsavebox{\SBoxBSetOnClickListener}
\codejmk[\small]{\SBoxBSetOnClickListener}{button.setOnClickListener($\ldots$)}

\JEDI{The app contains a bug, shown when calling \codej{execute}}
%
Our running example (code shown in \figref{feedremover-buggy-code}) is inspired by actual issues in AntennaPod~\cite{antennapodbug},
a podcast manager with 100,000+ installs,
and the Facebook SDK for Android~\cite{facebookbug}.
The essence of the issue is that a potentially time-consuming background task is started by a user interaction and implemented using the \codej{AsyncTask} framework class.
\Figref{feedremover-buggy-code} shows buggy code that can potentially violate the application-programming protocol for \codej{AsyncTask}.
The \usebox{\SBoxRemoverExecute} call (marked with \danger{}) throws an \codej{IllegalStateException} if the \codej{AsyncTask}~$t$ instance, pointed-to by \codej{remover}, is already running.
So a protocol rule for \codej{AsyncTask} is that \codej{$t$.execute()} cannot be called twice for the same \codej{AsyncTask}~$t$.
The \codej{IllegalStateException} type is commonly used to signal a protocol violation and has been shown to be a significant source of Android crashes~\cite{DBLP:conf/msr/KechagiaS14}.

In \figref{feedremover-buggy-code}, the \codej{RemoverActivity} defines an app window that, on creation (via the \codej{onCreate} callback), registers a click listener (via the \usebox{\SBoxBSetOnClickListener} call on \reftxt{line}{line-feedremover-register}).
%
This registration causes the framework to notify the app of a button click through the \codej{onClick} method.
When that happens, the \codej{onClick} callback starts the \codej{FeedRemover} asynchronous task (via the \usebox{\SBoxRemoverExecute} call on \reftxt{line}{line-feedremover-danger}).
What to do asynchronously is defined in the \codej{doInBackground} callback, and when the \codej{FeedRemover} task is done, the framework delegates to the \codej{onPostExecute} callback, which closes the \codej{RemoverActivity} (via the call to \usebox{\SBoxAFinish}).

We diagram a common-case execution trace in \figref{feedremover-trace-recorded}. Even though the app is buggy, the trace
does not witness the protocol violation. The exception does not manifest because the user only clicks once (\fmtevt{Click}) before the \codej{FeedRemover} task completes and generates the post-execute event (\fmtevt{PostExecute}). And so the \usebox{\SBoxExecute} callin on the \codej{AsyncTask} instance \codej{t} is executed only once before the activity is closed (cf. the \codej{onClick} and \codej{onPostExecute} callbacks in \figref{feedremover-buggy-code}).

\JEDI{An event-driven framework is ...}

If typically the \codej{Activity} is quickly destroyed after the button click, then seeing a protocol violation in a test is quite unlikely.
However, it is possible to click a second time before the \codej{AsyncTask} completes, thereby witnessing a protocol violation.
We show this error trace in
\figref{feedremover-trace-buggy}:
when the app invokes the callin \usebox{\SBoxExecute} for the second time in the second \fmtevt{Click} event, the framework is in a state that does not allow this transition.
We say that the callin invocation is \emph{disallowed} at this point, and apps must only invoke allowed callins.
While the original trace \fmtevt{Create};\fmtevt{Click};\fmtevt{PostExecute} does not concretely witness the protocol violation, it has sufficient information to predict the error trace \fmtevt{Create};\fmtevt{Click};\fmtevt{Click}.
It may, however, be difficult to reproduce this error trace:
the button must be pressed twice before the \usebox{\SBoxAFinish} method is called by the \fmtevt{PostExecute} event destroying the \codej{Activity}.
But how can we predict this error trace from the original one?

\newsavebox{\SBoxStatusPending}
\codejmk[\small]{\SBoxStatusPending}{remover.|ci:getStatus|() == AsyncTask.Status.PENDING}

\subsection{Predict Violations from Recorded Interactions}
\label{sec:overview-verification}

We define the dynamic lifestate verification problem as predicting an error trace that (possibly) witnesses a protocol violation from a trace of interactions or proving that no such error trace exists.
%
%
Concretely, the input to dynamic lifestate verification is an interaction trace like the one illustrated in
%
\figref{feedremover-trace-recorded}.
These traces record the sequence of invocations and returns of callbacks and callins between the framework and the app that result from an interaction sequence.
A recorded trace includes the concrete method arguments and return values (e.g., the instance \codej{t} from the diagrams corresponds to a concrete memory address).

\begin{figure}[!h]\centering
		\subcaptionbox{
			%
			A trace that does not witness a protocol violation since the callin
			\codej{(t:AsyncTask).execute()}
			on \codej{t} is executed only once.
			\label{fig:feedremover-trace-recorded}%
		}[0.5\linewidth]{
			\begin{minipage}[b][\heightof{\TheFeedRemoverTrace}]{1\linewidth}
				\TheFeedRemoverTrace
			\end{minipage}
		}
		\vspace{0.00mm}
		\subcaptionbox{
			The \fmtevt{Create}\rseq\fmtevt{Click}\rseq\fmtevt{Click} interaction sequences witnesses the no-\codej{execute}-call-on-already-executing-\codej{AsyncTask} protocol violation.
			\label{fig:feedremover-trace-buggy}
		}[0.48\linewidth]{
			\begin{minipage}[b][\heightof{\TheFeedRemoverTrace}]{1\linewidth}
				\TheFeedRemoverTraceBuggy
			\end{minipage}
		}
	\caption{We visualize the interface between an event-driven framework and an app as a dialog between two components.
		With execution time flowing downwards as a sequence events,
		control begins on the left with the framework receiving an event. Focusing on the
		\usebox{\SBoxHighlighted}
		\fmtevt{Click} event
		in \figref{feedremover-trace-recorded},
		when a user clicks on the button corresponding to object \codej{b} of type \codej{Button},
		the \codej{onClick} callback is invoked by the framework on the registered listener \codej{l}.
		For clarity,
		we write method invocations with type annotations (e.g., \usebox{\SBoxOnClick}), and variables \codej{b} and \codej{l} stand for some concrete instances (rather than program or symbolic variables).
		The app then delegates back to the framework by calling an API method to start an asynchronous task \codej{t} via \usebox{\SBoxExecute}.
		To connect with the app source code, we label the callins originating from the app timeline with the corresponding program point numbers in \figref{feedremover-buggy-code}.
		Here, we can see clearly a \emph{callback} as any app method that the framework can call (i.e., with an arrow to the right
		\raisebox{0.75ex}{\usebox{\SBoxCallbackArrow}}),
		and a \emph{callin} as any framework method that an app can call (i.e., with an arrow to the left
		\raisebox{0.75ex}{\usebox{\SBoxCallinArrow}}).
		We show returns with dashed arrows (but sometimes elide them when they are unimportant).
	}
\end{figure}

%
%
The main challenge, both for the app developer and dynamic lifestate verification, is that the relevant sequence of events that leads to a state where a callin is disallowed is \textit{hidden} inside the framework.
%
%
The developer must reason about the evolving internal state of the framework by considering the possible callback and callin interactions between the app and the framework to develop apps that both adhere to the protocol and behave intuitively.
To find a reasonable fix for the buggy app from \figref{feedremover-buggy-code}, let us consider again
the error trace shown in \figref{feedremover-trace-buggy}. Here, the developer has to reason that the \codej{(t:AsyncTask).execute()} callin is allowed as soon as \codej{t} is initialized by the call to \usebox{\SBoxOldTaskAlloc} in the \fmtevt{Create} event and is \emph{disallowed} just after the first call to \usebox{\SBoxExecute} in the first \fmtevt{Click} event.
%
That is, the developer must reason about
what sequence of events and callins determine when a callin is allowed or disallowed.
\JEDI{To determine if some app code, and hence callin, can be
executed, the developer must know when an event can happen
(enabled).}
Since callins are invoked inside callback methods and callback methods
are in turn invoked by the framework to notify the app of an event,
the internal framework state determines what events can happen when
and hence the callback control flow.
In particular, the internal framework state determines when the \fmtevt{Create} and \fmtevt{Click} events are \emph{enabled} (i.e., can happen) during the execution.
Thus to properly fix this app, the developer must ensure that \fmtevt{Create} happens before a \fmtevt{Click} and then only a single \fmtevt{Click} happens before a \fmtevt{PostExecute}. How can the app developer constrain the external interaction sequence to conform to this property?

\tikzstyle{safestate}=[state,draw=blue!50,fill=blue!10,minimum size=4mm]
\tikzstyle{errstate}=[state,draw=red!50,fill=red!10,minimum size=4mm]
\def\automatawidth{0.2\linewidth}
\newcommand{\FeedRemoverFixedCodeScale}{0.8}
\newcommand{\TheFeedRemoverTraceFixed}{\scalebox{0.65}{\usebox{\SBoxFeedRemoverTraceFixed}}}
\newcommand{\TheFeedRemoverFixedOne}{\scalebox{\FeedRemoverFixedCodeScale}{\usebox{\SBoxFeedRemoverFixedOne}}}
\newcommand{\AutomatonLifecycleRefined}{%
\scalebox{0.85}{\begin{tikzpicture}
\tikzstyle{every initial by arrow}=[initial text={}]
\tikzstyle{every loop}=[looseness=20]
\node[safestate,initial] (init) {};
\node[errstate] (created) [below of=init] {};
\path[->] (init) edge node [right] {\fmtevt{Create}} (created)
(init) edge [loop right] node {\fmtevt{Click}} (init)
(init) edge [in=90,out=120,loop] node [right] {\fmtevt{PostExecute}} (init)
%
(created) edge [loop right] node {\fmtevt{Click}} (created)
(created) edge [in=-120,out=-90,loop] node [below] {\fmtevt{PostExecute}} (created);
\end{tikzpicture}}%
}
\begin{figure}[!h]
\begin{tabular*}{\linewidth}{@{\extracolsep{\fill}}cc@{}}
\subcaptionbox{%
A \usebox{\SBoxSetEnabledFalse} call
prevents the user from clicking, 
triggering the \codej{onClick} callback.
\label{fig:feedremover-code-fixed}
}{
\begin{minipage}[b][\heightof{\TheFeedRemoverTraceFixed}][t]{0.35\linewidth}
\TheFeedRemoverFixedOne\vspace{-\medskipamount}
\par\scalebox{\FeedRemoverFixedCodeScale}{\usebox{\SBoxFeedRemoverFixedTwo}}\vspace{-\medskipamount}
\par\scalebox{\FeedRemoverFixedCodeScale}{\usebox{\SBoxFeedRemoverFixedThree}}\vspace{-\medskipamount}
\par\scalebox{\FeedRemoverFixedCodeScale}{\usebox{\SBoxFeedRemoverFixedFive}}
\end{minipage}
}
&
\subcaptionbox{%
The enabled callbacks and disallowed
callins are shown along the
\fmtevt{Create}\rseq\fmtevt{Click}\rseq\fmtevt{PostExecute} trace
from the fixed app in \protect\subref{fig:feedremover-code-fixed}.\label{fig:feedremover-trace-fixed}%
}{
\TheFeedRemoverTraceFixed
}
\end{tabular*}
\caption{A fixed version of the app from \figref{feedremover-buggy-code} that adheres to the application-programming protocol.
The annotations in \subref{fig:feedremover-trace-fixed} show that after the call to \codej{(b:Button).setEnabled(false)}, the \codej{l.onClick(b)} callback is no longer enabled, and thus the app can assume that the framework will not call \codej{l.onClick(b)} at this point.%
}
\end{figure}
\begin{figure}[!h]
\centering
\subcaptionbox{ 
{\color{red!50}False alarm} on the trivially sound, unconstrained, ``top'' abstraction.
\label{fig:feedremover-automata-top}
}{
\begin{minipage}[b][\heightof{\AutomatonLifecycleRefined}][t]{0.29 \linewidth}\centering
\scalebox{0.85}{\begin{tikzpicture}
	\tikzstyle{every initial by arrow}=[initial text={}]
	\tikzstyle{every loop}=[looseness=20]
	\node[errstate,initial] (init) {};
	\path[->] (init) edge [loop right] node {\fmtevt{Create}} (init)
	(init) edge [in=-60,out=-30,loop] node [right] {\fmtevt{Click}} (init)
	(init) edge [in=-120,out=-90,loop] node [below] {\fmtevt{PostExecute}} (init);
	\end{tikzpicture}}
\end{minipage}
} \hspace{0.2cm}
\subcaptionbox{ 
{\color{red!50}False alarm} on the \codej{Activity} lifecycle-refined abstraction.
\label{fig:feedremover-automata-lifecycle-noregistration}
}{
\begin{minipage}[b][\heightof{\AutomatonLifecycleRefined}][t]{0.29 \linewidth}\centering
\AutomatonLifecycleRefined
\end{minipage}
}\hspace{0.2cm}
\subcaptionbox{ 
{\color{red!50}False alarm} on the lifecycle with the \codej{Click} restricted to the active \codej{Activity} state (as shown in \figref{activitylifecycle}).
\label{fig:feedremover-automata-lifecycle}
}{
\begin{minipage}[b][\heightof{\AutomatonLifecycleRefined}][t]{0.29\linewidth}\centering
\scalebox{0.85}{\begin{tikzpicture}
	\tikzstyle{every initial by arrow}=[initial text={}]
	\tikzstyle{every loop}=[looseness=20]
	\node[safestate,initial] (init) {};
	\node[errstate] (created) [below of=init] {};
	\path[->] (init) edge node [right] {\fmtevt{Create}} (created)
	(init) edge [loop right] node {\fmtevt{PostExecute}} (init)
	(created) edge [loop right] node {\fmtevt{Click}} (created)
	(created) edge [in=-120,out=-90,loop] node [below] {\fmtevt{PostExecute}} (created);
	\end{tikzpicture}}
\end{minipage}
}
\subcaptionbox{ 
{\color{blue!50}Verified safe} when we consider the effect of \codej{Button.setEnabled($\text{\rm\ldots}$)}.
\label{fig:feedremover-automata-lifestate}
}{
\begin{minipage}[b][1.7cm][t]{0.4\linewidth}\centering
\scalebox{0.85}{\begin{tikzpicture}
	\tikzstyle{every initial by arrow}=[initial text={}]
	\tikzstyle{every loop}=[looseness=20]
	\node[safestate,initial,initial where=above] (init) {};
	\node[safestate] (created) [right=of init] {};
	\node[safestate] (clicked) [right=of created] {};
	\node[safestate] (executed) [below of=init] {};
	\path[->] (init) edge node [above] {\fmtevt{Create}} (created)
	(created) edge node [above] {\fmtevt{Click}} (clicked)
	(clicked) edge [bend left=10] node [below=5pt] {\fmtevt{PostExecute}} (executed);
	\end{tikzpicture}}
\end{minipage}
}\hspace{1cm}
\subcaptionbox{
This unsound abstraction is missing the \codej{PostExecute} edge.
\label{fig:feedremover-unsound-abstraction}
}{
\begin{minipage}[b][1.7cm][t]{0.4\linewidth}\centering
\scalebox{0.85}{
	\begin{tikzpicture}
	\tikzstyle{every initial by arrow}=[initial text={}]
	\tikzstyle{every loop}=[looseness=20]
	\node[safestate,initial,initial where=above] (init) {};
	\node[safestate] (created) [right=of init] {};
	\node[safestate] (clicked) [right=of created] {};
	\node[safestate] (executed) [below of=init] {};
	\path[->] (init) edge node [above] {\fmtevt{Create}} (created)
	(created) edge node [above] {\fmtevt{Click}} (clicked);
	\end{tikzpicture}}\\
\end{minipage}
}

\caption{
In previous works, models are generated for an application restricting the possible order of callbacks. In this figure, we show four sound abstractions with different levels of precision, indicating
whether they can verify our fixed application~\ref{fig:feedremover-code-fixed}, as well as one unsound abstraction.
}
\label{fig:feedremover-automata}

\end{figure}

In \figref{feedremover-code-fixed}, we show a fix based on the above insight that is particularly challenging to verify.
The fix adds line~\ref{line-feedremover-disable-fixed} that disables \codej{Button} \codej{button} to
indicate when the task has already been started.
Thus, this modified version does not violate the no-\codej{execute}-call-on-already-executing-\codej{AsyncTask} protocol
on \reftxt{line}{line-feedremover-danger-fixed}.
To reason precisely enough about this fix, we must know that the \usebox{\SBoxSetEnabledFalse} call changes internal framework state that prevents the \codej{onClick} from happening again.
%
%
%
Note that this need to reason about complex control flow arises from the interactions between just two framework types \codej{Button} and \codej{AsyncTask}---not to mention that these two are amongst the simplest framework types in Android. There is a clear need here for better automated reasoning tools to support the app developer.


%
\subparagraph{\toolname{} Approach.}
Our dynamic verification approach explores all the possible sequences of interactions that can be obtained by replicating, removing, and reordering the events in a trace.
By rearranging event traces, the algorithm statically explores different input sequences of events that a user interaction could generate.
The algorithm applied to the \fmtevt{Create}\rseq\fmtevt{Click}\rseq\fmtevt{PostExecute} trace in \figref{feedremover-trace-recorded} from the buggy app version indeed yields the error trace \fmtevt{Create}\rseq\fmtevt{Click}\rseq\fmtevt{Click} (shown in \figref{feedremover-trace-buggy}).
But more critically, our approach also makes it possible to prove that the fixed app version does not have any traces that violates the protocol (by
rearranging
\fmtevt{Create}\rseq\fmtevt{Click}\rseq\fmtevt{PostExecute}).

\JEDI{The algorithm uses the set of enabled states and allowed callin
when searching for execution traces that violate the protcol}
Central to our approach is 
capturing
the essential, hidden framework state---tracking the set of enabled callbacks and the set of disallowed callins.
\Figref{feedremover-trace-fixed} illustrates this model state along a trace
from the fixed app. After the first \fmtevt{Click}, the application disables the button to prevent a second \fmtevt{Click} via the call to \usebox{\SBoxSetEnabledFalseCall}, which at that point removes \codej{l.onClick(b)} from the set of enabled callbacks the framework can trigger.

\JEDI{The dynamic verification algorithm is solved by a reduction to
a model checking problem.}
\toolname{} addresses the dynamic verification problem by reducing it to a model checking problem.
%
The model is a transition system with
\begin{inparaenum}[(i)]%
\item states abstracting the set of enabled callbacks and disallowed callins and
\item transitions capturing the possible replication, removing, and reordering of a given interaction trace.
\end{inparaenum}
The safety property of interest is that the transition system never
visits a disallowed callin.
%
%
\JEDI{Main issue, the protocol is hidden, so it must be specified}%
%
How can we construct such a transition system that over-approximates concrete behavior while being precise enough to make alarm triage feasible?
As alluded to in \secref{introduction}, lifestate specification is crucial here.

\subsection{Specify Event-Driven Protocols and Model
Callback Control Flow}
\label{sec:overview-specification}

\JEDI{We define the lifestate language to specify the model as a dialog between the application and framework.}

In \figref{feedremover-automata}, we illustrate the essence of callback control-flow modeling
as finite-state automata that over-approximate rearrangements of the \fmtevt{Create}\rseq\fmtevt{Click}\rseq\fmtevt{PostExecute} trace shown in
\figref{feedremover-trace-fixed}.
Automaton~\ref{fig:feedremover-automata-top} 
exhibits the trivially sound, unconstrained, ``top'' abstraction that considers all replications, removals, and reorderings of the interaction trace.
This abstraction is the one that assumes all callbacks are always enabled.
Since a possible trace in this abstraction includes two \fmtevt{Click} events, a sound verifier must alarm.
Meanwhile, Automaton~\ref{fig:feedremover-automata-lifecycle-noregistration} shows a refined abstraction encoding the Android-specific \codej{Activity} lifecycle.
The abstraction is framework-specific but application-independent and captures that the \fmtevt{Create} event cannot happen more than once.
The abstraction shown by Automaton~\ref{fig:feedremover-automata-lifecycle-noregistration} is also insufficient to verify the trace from the fixed app because two \fmtevt{Click} events are still possible.
%

%

\JEDI{Additional ad-hoc rules that enrich the non-lifecycle constraints are also not enough}
Automaton~\ref{fig:feedremover-automata-lifecycle} shows a refined, \lifecyclepp{} abstraction that considers the \codej{Activity} lifecycle with additional constraints on an ``attached'' \fmtevt{Click} event.
This abstraction is representative of the current practice in callback control-flow models
(e.g.,
\cite{DBLP:conf/pldi/ArztRFBBKTOM14,DBLP:conf/pldi/MaiyaKM14,DBLP:conf/oopsla/BlackshearCS15}).
While Automaton~\ref{fig:feedremover-automata-lifecycle} restricts the \fmtevt{Click} event to come only after the \fmtevt{Create} event, the abstraction is still too over-approximate to verify that the trace 
from the fixed app is safe---two \fmtevt{Click} events are still possible with this model.
But worse is that this model is still, in essence, a lifecycle model that is constrained by Android-specific notions like \codej{View} attachment, \codej{Listener} registration, and the ``live'' portion of lifecycles. In existing analysis tools, such constrained lifecycle models are typically hard-coded into the analyzer.


\newcommand{\pa}[1][]{\MVarNamed{\specvar}{#1}}
\newsavebox{\SBoxSetEnabledFalseSpec}
\codejmk[\footnotesize]{\SBoxSetEnabledFalseSpec}{($\pa[b]$:Button).setEnabled(false)}
\newsavebox{\SBoxOnClickSpec}
\codejmk[\footnotesize]{\SBoxOnClickSpec}{($\pa[l]$:OnClickListener).onClick($\pa[b]$:Button)}
\newsavebox{\SBoxExecuteSpec}
\codejmk[\footnotesize]{\SBoxExecuteSpec}{($\pa[t]$:AsyncTask).execute()}
\newsavebox{\SBoxOnPostExecuteSpec}
\codejmk[\footnotesize]{\SBoxOnPostExecuteSpec}{($\pa[t]$:AsyncTask).onPostExecute()}

We need a better way to capture how the application may affect callback control flow.
%
In this example, we need to capture the effect of the callin
\usebox{\SBoxSetEnabledFalse} at
\reftxt{line}{line-feedremover-disable-fixed} in
\figref{feedremover-code-fixed}, which is the only difference with
the buggy version in \figref{feedremover-buggy-code}.
The modeling needs to be expressive to remove such infeasible traces and compositional to express state changes independently.
Thus, the role of
lifestate specification is to describe how the internal model state
is updated by observing the history of
intertwined callback and callin invocations. For example, we write
\[\begin{tabular*}{\linewidth}{@{}L@{}}
\usebox{\SBoxSetEnabledFalseSpec} \rdisable \usebox{\SBoxOnClickSpec}\;\text{(for all \pa[l], \pa[b])}
\end{tabular*}
\]
to model when \usebox{\SBoxSetEnabledFalseSpec} is invoked,
the click callback 
is \emph{disabled}
on the same button~\pa[b] (on all listeners~\pa[l]).
Also, we similarly specify the safety property of interest
\[\begin{tabular*}{\linewidth}{@{}L@{}}
\usebox{\SBoxExecuteSpec} \rdisable \usebox{\SBoxExecuteSpec}\;\text{(for all \pa[t])}
\end{tabular*}\]
that when \usebox{\SBoxExecuteSpec} is called on a task~\pa[t], it
\emph{disallows} itself. And analogously, lifestates include specification forms
for \emph{enabling callbacks} or \emph{allowing callins}.

%

%
%

\JEDI{Lifestate are a uniform language to model callback control-flow and specify event-driven programming protocols}
Lifestate uniformly models the callback control-flow and specifies event-driven application-programming protocols.
The rules that enable and disable callbacks model what callbacks the framework can invoke at a specific point in the execution of the application, while
the rules that disallow and allow callins specify what callins the application must invoke to respect the protocol.
%
What makes lifestate unique compared to
typestates~\cite{strom+1986:typestate:-programming} or lifecycle automata is
this unification of the intertwined effects of callins and callbacks on each
other.


\JEDI{Very complex, so it is also ``easier'' to mess up. How can we identify errors in a lifestate specification?}

The complexity of the implicit callback control flow is what makes expressing and writing correct models challenging. An issue whose importance is often under-estimated when developing callback control-flow models is how much the model faithfully reflects the framework semantics. How can we \emph{validate} that a lifestate specification is a correct model of the event-driven framework?

\subparagraph{Validating Event-Driven Programming Protocols.}

\JEDI{The correct lifestate-abstraction contain the postexecute edge and this is specified with an enable lifestate rule}
As argued in \secref{introduction}, a key concern when developing a framework model is that it must over-approximate the possible real behavior of the application. The ``top'' model as shown in Automaton~\ref{fig:feedremover-automata-top} trivially satisfies this property, and it may be reasonable to validate an application-independent lifecycle model like Automaton~\ref{fig:feedremover-automata-lifecycle}.
However, as we have seen, verifying correct usage of event-driven protocols typically requires callback control-flow models with significantly more precision.

Automaton~\ref{fig:feedremover-automata-lifestate} shows a correct lifestate-abstraction that contains an edge labeled \codej{PostExecute}. We express this edge with the
rule shown below:
\[\begin{tabular*}{\linewidth}{@{}L@{}}
\usebox{\SBoxExecuteSpec} \renable \usebox{\SBoxOnPostExecuteSpec}\;\text{(for all \pa[t])}
\end{tabular*}\]
This rule states that when \usebox{\SBoxExecuteSpec} is called, its effect is to enable the callback \usebox{\SBoxOnPostExecuteSpec} on the \codej{AsyncTask}~\pa[t].

\JEDI{If we do not add the lifestate rule, we obtain a wrong abstraction that misses a key transition.}

%
If we do not model
this
rule, we obtain the abstraction in Automaton~\ref{fig:feedremover-unsound-abstraction}.
The lifestate model is \emph{unsound} since it misses the \codej{PostExecute} edge.

\JEDI{We validate the lifestate model using the trace eventually exposing an unsound behavior.}

The trace \fmtevt{Create}\rseq\fmtevt{Click}\rseq\fmtevt{PostExecute} shown in
\figref{feedremover-trace-fixed}
is a witness of the unsoundness of the abstraction:
Automaton~\ref{fig:feedremover-unsound-abstraction} accepts only proper prefixes of the trace (e.g., \fmtevt{Create}\rseq\fmtevt{Click}), and hence the abstraction does not capture all the possible traces of the app.
We can thus use interaction traces to validate lifestate rules: a set of lifestate rules is valid if the abstraction accepts all the interaction traces.
The validation applied to the abstraction shown in Automaton~\ref{fig:feedremover-unsound-abstraction} demonstrates that the abstraction accepts \fmtevt{Create}\rseq\fmtevt{Click} as the longest prefix of the trace \fmtevt{Create}\rseq\fmtevt{Click}\rseq\fmtevt{PostExecute}. This information helps to localize the cause for unsoundness since we know that after the sequence \fmtevt{Create}\rseq\fmtevt{Click}, the callback \fmtevt{PostExecute} is (erroneously) disabled.
%

The encoding of the abstraction from  lifestate rules is a central step to perform model validation and dynamic verification.
At this point, we still cannot directly encode the abstraction since the lifestate rules contain universally-quantified variables.
How can we encode the lifestate abstraction as a transition system amenable to check language inclusion for validation, and to check safety properties for dynamic verification?





\subparagraph{From Specification to Validation and Verification.}

Generalizing slightly, we use the term \emph{message} to refer to any observable
interaction between the framework and the app.
Messages consist of
invocations to and returns from callbacks and callins. The abstract state
of the transition system is then a pair consisting of the
\emph{permitted-back messages} from framework to app and the \emph{prohibited-in
messages} from app to framework. And thus generalizing the example
rules shown above, a lifestate specification is a set of rules whose meaning is,
\[
\parbox{0.9\linewidth}{%
\emph{If} the message history \emph{matches} \rregexp{}, \emph{then} the abstract state is updated according to the specified \emph{effect} on the set of permitted-back and prohibited-in messages.%
}%
\]
There are many possible choices and tradeoffs for the matching language \rregexp{}. As is
common, we consider a regular ex\-press\-ion-based (i.e., finite automata-based)
matching language.

\JEDI{The quantified variables are hard to encode, hence we ``finitize'' them by using the trace}
We exploit the structure of the validation and dynamic verification problem to encode the lifestate abstraction.
In both problems, the set of possible objects and parameters is finite and determined by the messages recorded in the trace. We exploit this property to obtain a set of \emph{ground} rules (rules without variables).
%
%
%
%
\JEDI{Since the ground rules are regular expressions, they are
easily encoded in the model used for verification}%
We can then encode each ground rule in a transition system.
Since the rule is ground, the encoding is standard: each regular expression is converted to an automaton and then encoded in the transition system, changing the permitted-prohibited state as soon as the transition system visits a trace accepted by the regular expression, which implicitly yields a model like automata~\ref{fig:feedremover-automata-lifestate}.

Lifestate offers a general and flexible way to specify the possible future messages in terms of observing the past history of messages. It, however, essentially leaves the definition of messages and what is observable abstract. What observables characterize the interactions between an event-driven framework and an app that interfaces with it? And how do these observables define event-driven application-programming protocols and callback control flow?

\subsection{Event-Driven App-Framework Interfaces}
\label{sec:overview-concrete}

Lifestate rules are agnostic to the kinds of messages they match and effects they capture on the internal abstract state.
To give meaning to lifestates, we formalize the essential aspects of the app-framework interface in an abstract machine model called $\semname$ in \secref{concrete}.
This abstract machine model formally characterizes what we consider an \emph{event-driven framework}.
The $\semname$ abstract machine crisply defines the messages that the app and the framework code exchange and a formal correspondence between concrete executions of the program
and the app-framework interface.
We use this formal correspondence to define the semantics of the lifestate framework model, its validation problem, and protocol verification.

We do not intend for $\semname$ to capture all aspects of something as complex as Android; rather, the purpose of $\semname$ is to define a ``contract'' by which to consider a concrete event-driven framework implementation. And thus, $\semname$ also defines the dynamic-analysis instrumentation we perform to record observable traces from Android applications that we then input to the \toolname{} tool to either validate a specification or verify protocol violations.

\subparagraph{Preview.} We have given a top-down overview of our approach, motivating with the dynamic protocol verification problem the need for having both a precise callback control-flow model and an event-driven protocol specification. We also presented how the lifestate language addresses this need capturing the intertwined effect of callins and callbacks.
In the next sections, we detail our approach in a bottom-up manner---beginning
with formalizing the $\semname$ abstract machine model.
We show that, assuming such a model of execution, it is possible to provide a sound abstraction of the framework (i.e., no real behavior of the framework is missed by the abstraction) expressed with a lifestate model.
We then formalize how we validate such models and how we use lifestates to verify the absence of protocol violations.

\section{Defining Event-Driven Application-Programming Protocols}
\label{sec:concrete}

Following \secref{overview}, we want to capture the essence of the
app-framework interface with respect to framework-imposed programming protocols.
\JEDI{\semname formalize an event-driven application}%
To do so, we first formalize a small-step operational semantics for event-driven
programs with an abstract machine model \semname.
The \semname{} abstract machine draws on standard techniques but explicitly highlights enabled events and disallowed
callins to precisely define event-driven protocols.
\JEDI{\semname formalizes what is observable in the framework }%
We then instrument this semantics to formalize the interface
of the event-driven framework with an app, thereby defining
the traces of the observable app-framework interface of a \semname program.
%
%


This language is intentionally minimalistic to center on capturing just the interface between event-driven frameworks and their client applications. By design, we leave out many aspects of real-world event-driven framework implementations (e.g., Android, Swing, or Node.js), such as typing, object-orientation, and module systems that are not needed for formalizing the dialogue between frameworks and their apps (cf. \secref{overview}). Our intent is to illustrate, through examples, that event-driven frameworks could be implemented in $\semname{}$ and that $\semname{}$ makes explicit the app-framework interface to define \emph{observable traces} consisting of \emph{back-messages} and \emph{in-messages} (\secref{instrumented-semantics}).

\begin{figure}[t]\centering
\subcaptionbox{The syntax and the semantic domains.\label{fig:model-syntax}}{
\begin{minipage}{\linewidth}
\small
\begin{mathpar}
\begin{grammar}[@{}r][l@{}]
expressions &
\expr \in \ExprSet
& \bnfdef & \enBind{\val_1}{\val_2}
\bnfalt \enInvoke{\val}
\bnfalt \enDisallow{\val}
\bnfalt \enAllow{\val}
& thunks and calls
\\
&& \bnfalt & \enEnable{\val} \bnfalt \enDisable{\val}
\bnfalt \enForce{\thunk}
& events and forcing
\\
&& \bnfalt &
\val
\bnfalt \enLet{\var}{\expr_1}{\expr_2}
\bnfalt \cdots
& other expressions
\\
functions &
\fun & \bnfdef & \enFun[\pkg]{\var}{\expr}
\\
packages &
\pkg & \bnfdef & \enApp \bnfalt \enFwk
\\
values &
\val \in \ValSet & \bnfdef & \var \bnfalt \enClosure{\fun}{\env} \bnfalt \thunk \bnfalt \enSkip \bnfalt \cdots \bnfalt \enMe
\end{grammar}
\end{mathpar}
\begin{mathpar}
\text{variables} \quad \var \in \VarSet
\and
\text{thunks} \quad \thunk \in \ThunkSet \bnfdef \enThunk{\enClosure{\fun}{\env}}{\val}
\and
\text{thunk stores} \quad
\eventmap, \callinmap \bnfdef \cdot \bnfalt \eventmap \extMsg{\thunk}
\end{mathpar}
\begin{mathpar}
\text{continuations} \quad
\cont
\bnfdef \enSkipK \bnfalt \enLetK{\var}{\cont}{\expr}{\env}
\bnfalt \thunk \bnfalt \enFrame{\thunk}{\cont}
\and
\text{states} \quad
\stateconc \in \StateSet \bnfdef \enState \bnfalt \enDisallowed
\end{mathpar}
\end{minipage}
}
\subcaptionbox{Semantics. Explicitly enable, disable, disallow, and allow thunks.\label{fig:model-semantics-event}}{
\begin{minipage}{0.98\linewidth}
\hfill\fbox{$\jstep{\stateconc}{\stateconc'}$}
\footnotesize\begin{mathpar}%
\inferrule[Enable]{
}{
	\jstep{ \enState[\enEnable{\thunk}] }%
	{ \enState[ \thunk ][ \env ][ \store ][ \eventmap\extMsg{\thunk} ] }
}
\inferrule[Disable]{
}{
	\jstep{ \enState[\enDisable{\thunk}][\env][\store][ \eventmap{\extMsg{\thunk}} ] }%
	{ \enState[ \thunk ] }
}
\inferrule[Event]{
	\thunk \in \eventmap
}{
	\jstep{ \enState[ \val ][ \env ][ \store ][ \eventmap ][ \callinmap ][ \enSkipK ] }%
	{ \enState[ \enForce{\thunk} ][ \env ][ \store ][ \eventmap ][ \callinmap ][ \thunk ] }%
}
\\
\inferrule[Disallow]{
}{
	\jstep{ \enState[\enDisallow{\thunk}] }%
	{ \enState[ \thunk ][ \env ][ \store ][ \eventmap ][ \callinmap\extMsg{\thunk} ] }
}

\inferrule[Allow]{
}{
	\jstep{ \enState[\enAllow{\thunk}][\env][\store][ \eventmap ][ \callinmap\extMsg{\thunk} ] }%
	{ \enState[ \thunk ] }
}
\\
\inferrule[Invoke]{
	\thunk \notin \callinmap
}{
	\jstep{ \enState[ \enInvoke{\thunk} ] }%
	{ \enState[ \enForce{\thunk} ] }
}
\;
\inferrule[InvokeDisallowed]{
  \thunk \in \callinmap
}{
  \jstep%
        { \enState[ \enInvoke{\thunk} ] }%
        { \enDisallowed }
}
\;
\inferrule[Bind]{
}{
	\jstep{ \enState[\enBind{\fun}{\val}] }%
	{ \enState[ \enThunk{\enClosure{\fun}{\env_1}}{\val} ] }
}
\\
\inferrule[Force]{
\enThunk{\enClosure{ (\enFun[\pkg']{\var'}{\expr'}) }{\env'}}{\val'} = \thunk
}{
  \jstep{ \enState[ \enForce{\thunk} ] }%
        { \enState[ \subst{\thunk}{\enMe}\subst{\val'}{\var'}\expr' ][\env][\store][\eventmap][\callinmap]%
        [ \enFrame{\thunk}{\cont} ] }
}
\inferrule[Return]{ }{
	\jstep{ \enState[ \val ][ \env ][ \store ][ \eventmap ][ \callinmap ][ \enFrame{\thunk}{\cont} ] }%
	{ \enState[ \val ][ \env ][ \store ][ \eventmap ][ \callinmap ][ \cont ] }%
}
\inferrule[Finish]{ }{
	\jstep{ \enState[ \val ][ \env ][ \store ][ \eventmap ][ \callinmap ][ \thunk ] }%
	{ \enState[ \val ][ \env ][ \store ][ \eventmap ][ \callinmap ][ \enSkipK ] }%
}
\\
\inferrule[Let]{ }{
  \jstep{ \enState[ \enLet{\var}{\expr_1}{\expr_2} ] }%
        { \enState[ \expr_1 ][ \env ][ \store ][ \eventmap ][ \callinmap ][ \enLetK{\var}{\cont}{\expr_2}{\env} ] }%
}

\inferrule[Continue]{ }{
  \jstep{ \enState[ \val ][ \env ][ \store ][ \eventmap ][ \callinmap ][ \enLetK{\var}{\cont}{\expr_2}{\env} ] }%
        { \enState[ \subst{\val}{\var}\expr_2 ] }%
}
\end{mathpar}
\end{minipage}
}
\caption{\semname, a core model of
  event-driven programs capturing \emph{enabledness} of events and
  \emph{disallowedness} of invocations.}
\label{fig:model-syntax-semantics}
\end{figure}
\newsavebox{\SBoxEnPostExecuteInvoke}
\begin{lrbox}{\SBoxEnPostExecuteInvoke}\small
\begin{lstlisting}[language=Enable,alsolanguage=exthighlighting]
let cb = bind |cb:onPostExecute| t in invoke cb
\end{lstlisting}
\end{lrbox}
\newsavebox{\SBoxEnPostExecuteEvtEnable}
\begin{lrbox}{\SBoxEnPostExecuteEvtEnable}\small
\begin{lstlisting}[language=Enable,alsolanguage=exthighlighting]
let h = bind handlePostExecute t in enable h
\end{lstlisting}
\end{lrbox}
\newsavebox{\SBoxEnPostExecute}
\begin{lrbox}{\SBoxEnPostExecute}\small
\begin{lstlisting}[language=Enable,alsolanguage=exthighlighting]
let |cb:onPostExecute| = (t =>${}_{\enApp}$ $\ldots$) in
let handlePostExecute = (t =>${}_{\enFwk}$ let cb = bind |cb:onPostExecute| t in invoke cb) in
\end{lstlisting}
\end{lrbox}
\newsavebox{\SBoxEnExecute}
\begin{lrbox}{\SBoxEnExecute}\small
\begin{lstlisting}[language=Enable,alsolanguage=exthighlighting]
let |ci:execute| = (t =>${}_{\enFwk}$ disallow thk; $\ldots$ let h = bind handlePostExecute t in enable h)
\end{lstlisting}
\end{lrbox}

\subsection{Syntax: Enabling, Disabling, Allowing, and Disallowing}
\label{sec:thunks}

The syntax of \semname is shown at the top of \figref{model-syntax},
which is a $\lambda$-calculus in a let-normal form.
The first two cases of expressions $\expr$ split the standard call-by-value function
application into multiple steps
(similar to call-by-push-value~\cite{DBLP:journals/lisp/Levy06}).
The \enBind{\fun}{\val}
expression creates a thunk $\thunk = \enThunk{\fun}{\val}$ by binding a function value
$\enClosure{\fun}{\env}$ with an argument value $\val$.
We abuse notation slightly by using $\fun$ as the meta-variable for function values
(rather than as a terminal symbol).
A thunk may be forced by direct invocation $\enInvoke{\thunk}$---or
indirectly via event dispatch.
\begin{example}[Applying a Function] Let \codej{t} be bound to an \codej{AsyncTask} and \codej{onPostExecute} to an app-defined callback (e.g., \codej{onPostExecute} from \figref{feedremover-code-fixed}), then the direct invocation of a callback from the framework can be modeled by the two steps of binding and then invoking:
\par\noindent\begin{minipage}{\linewidth}\centering
\usebox{\SBoxEnPostExecuteInvoke}
\end{minipage}
\end{example}
%
%
Now in $\semname$,
a thunk $\thunk$ may or may not have the \emph{permission} to be
forced.
Revoking and re-granting the permission to force a thunk via direct invocation is captured by the expressions
\enDisallow{\thunk} and \enAllow{\thunk}, respectively. A protocol
violation can thus be modeled by an application invoking a
\emph{disallowed} thunk.
%

%
The direct invocation expressions are mirrored with expressions for
event dispatch. An \enEnable{\thunk} expression \emph{enables} a thunk
$\thunk$ for the external event-processing system (i.e., gives the system permission to force the thunk \thunk),
while the \enDisable{\thunk} expression \emph{disables} the thunk
$\thunk$.
%
%
%
\begin{example}[Enabling an Event] Let \codej{t} be bound to an \codej{AsyncTask} and \codej{handlePostExecute} to an internal framework-defined function for handling a post-execute event, then enqueuing such an event can be modeled by the two steps of binding then enabling:
\par\noindent\begin{minipage}[b][0.25cm][t]{\linewidth}\centering
\usebox{\SBoxEnPostExecuteEvtEnable}
\end{minipage}
\end{example}

By separating function application and event dispatch into binding to create a thunk $\thunk = \enThunk{\fun}{\val}$ and then forcing it, we uniformly make thunks the value form that can be granted permission to be invoked (via \enAllow{\thunk}) or for event dispatch (via \enEnable{\thunk}).
The \enForce{\thunk} expression is then an intermediate
that represents a thunk that is forcible (i.e., has been permitted
for forcing via \enAllow{\thunk} or \enEnable{\thunk}).


The remainder of the syntax is the standard part of the language: values $\val$,
variable binding $\enLet{\var}{\expr_1}{\expr_2}$, and whatever
other operations of interest $\cdots$ (e.g., arithmetic, tuples, control flow,
heap manipulation). That is, we have made explicit the expressions to expose the app-framework interface and can imagine whatever standard language features in $\cdots$ in framework implementations.
The values $\val$ of this expression language are variables $\var$,
function values $\enClosure{\fun}{\env}$,
thunks $\thunk$, unit $\enSkip$,
and whatever other base values of interest $\cdots$.
Two exceptions are that
(1) the currently active thunk is available via the $\enMe$ identifier
(see \secref{lambdalife-semantics}) and
(2) functions
$\enFun[\pkg]{\var}{\expr}$ are tagged with a package
$\pkg$ (see \secref{instrumented-semantics}).
\subsection{Semantics: Protocol Violations}
\label{sec:lambdalife-semantics}

At the bottom of \figref{model-syntax},
we consider an abstract machine model enriched
with an \emph{enabled-events} store $\eventmap$,
and a \emph{disallowed-calls} store $\callinmap$.
These are finite sets
of thunks, which we write as a list $\thunk_1\extMsg{\cdots \extMsg{\thunk_n}}$.
The enabled-events store $\eventmap$ saves thunks that are permitted to
be forced by the event loop, while the disallowed-calls store $\callinmap$ lists thunks that
are \emph{not} permitted to be forced by invocation.
These thunk stores make explicit the event-driven application-programming protocol (that might otherwise be implicit in, for example, flag fields and conditional guards).

A machine state
$\stateconc\colon \enState$ consists of an expression $\expr$, enabled
events $\eventmap$, disallowed calls $\callinmap$, and
a continuation $\cont$.
A continuation $\cont$ can be the top-level continuation $\enSkipK$ or
a continuation for returning to the body of a $\enkwLet$ expression,
which are standard.
Continuations are also used to record the active thunk via
$\thunk$ and $\enFrame{\thunk}{\cont}$ corresponding to the
run-time stack of activation records.
These continuation forms record the active thunk and are for defining messages and
the app-framework interface in \secref{instrumented-semantics}.
Since events occur non-deterministically and return to the main
event loop, it is reasonable to assume that a state $\stateconc$ should also include a
heap, and the expression language should have heap-manipulating
operations through which events communicate.
We do not, however, formalize heap operations since they are standard.
%

We define an operational semantics in terms of the judgment form
$\jstep{\stateconc}{\stateconc'}$ for a small-step transition
relation.
In \figref{model-semantics-event}, we show the inference
rules defining the reduction steps related to
enabling-disabling,
disallowing-allowing,
invoking, creating,
and finally forcing
thunks. The rules follow closely the informal semantics discussed in
\secref{thunks}.
Observe that \TirName{Enable} and \TirName{Allow} both
permit a thunk to be forced, and \TirName{Disable} and
\TirName{Disallow} remove the permission to be forced for a thunk.
The
difference between \TirName{Enable} and \TirName{Disable} versus
\TirName{Allow} and \TirName{Disallow} is that the former pair
modifies the enabled events $\eventmap$, while the latter
touches
the disallowed calls $\callinmap$.
%
%

The \TirName{Event} rule says that when the expression is a value
$\val$ and the continuation is the top-level continuation $\enSkipK$,
then a thunk is non-deterministically chosen from the enabled
events $\eventmap$ to force.
Observe that an enabled event remains enabled after an \TirName{Event}
reduction, hence \semname can model both
events that do not self-disable (e.g., the \codej{Click} event from \secref{overview}) and
those that are self-disabling (e.g., the \codej{Create} event).
The \TirName{Invoke} rule has a similar effect, but it checks that the
given thunk is not disallowed in $\callinmap$ before forcing.
The \TirName{InvokeDisallowed} rule states that a disallowed thunk
terminates the program in the $\enDisallowed$ state.
And the \TirName{Bind} rule simply states that thunks are created by binding an actual argument to a function value.

The \TirName{Force} rule implements the ``actual application'' that
reduces to the function body $\expr'$ with the argument $\val'$
substituted for the formal $\var'$ and the thunk substituted for the identifier $\enMe$, that is, $\subst{\thunk}{\enMe}\subst{\val'}{\var'}\expr'$.
To record the stack of activations, we push the forced thunk
$\thunk$ on the continuation (via $\enFrame{\thunk}{\cont}$).
The \TirName{Return} and \TirName{Finish} rules simply state that the
recorded thunk $\thunk$ frames are popped on return from a
\TirName{Force} and \TirName{Event}, respectively.
The \TirName{Return} rule returns to the caller via the continuation $\cont$, while the \TirName{Finish} rule returns to the top-level event loop $\enSkipK$.
The last line with the \TirName{Let} and \TirName{Continue} rules
describe, in a standard way, evaluating let-binding.

A program $\expr$ violates the event-driven protocol if it ends in the $\enDisallowed$ state from the initial state
$\enState[ \expr ][ \cdot ][ \cdot ][ \cdot ][ \cdot ][ \enSkipK ]$. 
\begin{example}[Asserting a Protocol Property.] The no-\codej{execute}-call-on-already-executing-\codej{AsyncTask} protocol can be captured by a $\enkwDisallow$. We let \codej{execute} be a framework function (i.e., tagged with $\enFwk$) that takes an \codej{AsyncTask}  \codej{t}.
%
\par\noindent\begin{minipage}[b][0.5cm][t]{\linewidth}\centering
\usebox{\SBoxEnExecute}
\vspace{\belowdisplayskip}
\end{minipage}
\par\noindent
The \codej{execute} function first disallows itself (via \enDisallow{\enMe}) and does some work  (via $\ldots$) before enabling the \codej{handlePostExecute} event handler (writing $\expr_1; \expr_2$ as syntactic sugar for sequencing). The \enDisallow{\enMe} asserts that this thunk cannot be forced again---doing so would result in a protocol violation (i.e., the $\enDisallowed$ state).
\end{example}
%


In contrast to an event-driven framework implementation, the state of a $\semname$
program does not have a queue. As we see here, a queue is an implementation detail
not relevant for capturing event-driven programming protocols.
Instead, $\semname$ models the external environment, such as, user
interactions, by the non-deterministic selection of an enabled event.


\subsection{Messages, Observable Traces, and the App-Framework Interface}
\label{sec:instrumented-semantics}

To minimally capture how a program is composed of separate framework and app code,
we add some simple syntactic restrictions to $\semname$ programs.
%
%
%
%
Function values $\fun$ tagged with the $\enFwk$ are framework code and the $\enApp$ tag labels app code.
%
We express a framework implementation
$\FwkImpl$ with a finite set of framework functions
$\FunFwkSet$ and an initialization function $\funinit \in \FunFwkSet$.
A program $\expr$ uses the framework implementation
if it first invokes the function $\funinit$,
and all the functions labeled as
$\enFwk$ in $\expr$ are from $\FunFwkSet$.

In a typical, real-world framework implementation, the framework implicitly defines the application-programming protocol with internal state to check for protocol violations.
The \TirName{Enable}, \TirName{Disable}, \TirName{Allow}, and \TirName{Disallow} transitions make explicit the event-driven protocol specification in $\semname$.
Thus, it is straightforward to capture that framework-defined protocols by syntactically prohibiting the app from using $\enEnable{\thunk}$, $\enDisable{\thunk}$, $\enAllow{\thunk}$, and $\enDisallow{\thunk}$. Again, the enabled-event store $\eventmap$ and the disallowed-call store $\callinmap$ in $\semname$ can be seen as making explicit the implicit internal state of event-driven frameworks that define their application-programming protocols.


The app interacts with the framework only by ``exchanging messages.''
The app-framework dialogue diagrams
from Figures~\ref{fig:feedremover-trace-recorded},~\ref{fig:feedremover-trace-buggy},
and~\ref{fig:feedremover-trace-fixed} depicts the notion of messages
as arrows back-and-forth between the framework and the app.
The framework invokes callbacks and returns from callins (the arrows from left to right), while the app invokes callins and returns from callbacks (the arrows from right to left).
To formalize this dialogue,
we label the observable transitions in the judgment form and define an
\emph{observable trace}---a trace formed only by these observable messages.
Being internal to the framework, the \TirName{Enable}, \TirName{Disable}, \TirName{Allow}, and \TirName{Disallow} transitions are hidden, or unobservable, to the app.

\begin{figure}[tb]\small
\begin{mathpar}[\lineskip=0.2em]
\text{back-messages} \quad \glabelFwk{} \in \GMsgSetFwk \bnfdef \enCb{\thunk} \bnfalt \enCiRet{\val}{\thunk}
\and
\text{in-messages} \quad \glabelApp{} \in \GMsgSetApp \bnfdef \enCi{\thunk} \bnfalt \enCbRet{\val}{\thunk}
\\
\text{messages} \quad \glabel{} \in \GMsgSet
\bnfdef \glabelFwk{} \bnfalt \glabelApp{} \bnfalt \enDis{\glabelApp{}} \bnfalt \emptyword
\and
\text{observable traces} \quad \mword{} \in \seqof{\TransSet} \bnfdef \emptyword \bnfalt \mword{}\glabel{}
\end{mathpar}
%
\fbox{$\jstepins{\stateconc}{\stateconc'}$}
\footnotesize\begin{mathpar}
\inferrule[ForceCallback]{
\enThunk{\enClosure{ (\enFun[\enApp]{\var'}{\expr'}) }{\env'}}{\val'} = \thunk
  \\
  \enFwk = \Package(\cont)
}{
  \jstepins[\enCb{\thunk}]
        { \enState[ \enForce{\thunk} ] }%
        { \enState[ \subst{\thunk}{\enMe}\subst{\val'}{\var'}\expr']%
[ \env ]%
[ \store ]%
[ \eventmap ][ \callinmap ][ \enFrame{\thunk}{\cont} ] }
}

\inferrule[ForceCallin]{
  \enThunk{\enClosure{ (\enFun[\enFwk]{\var'}{\expr'}) }{\env'}}{\val'} = \thunk
  \\
  \enApp = \Package(\cont)
}{
  \jstepins[\enCi{\thunk}]
        { \enState[ \enForce{\thunk} ] }%
        { \enState[ \subst{\thunk}{\enMe}\subst{\val'}{\var'}\expr'][]%
[ \store ]%
[ \eventmap ][ \callinmap ][ \enFrame{\thunk}{\cont} ] }
}
\\
\inferrule[ReturnCallin]{
  \enThunk{\enClosure{ (\enFun[\enFwk]{\var'}{\expr'}) }{\env'}}{\val'} = \thunk
  \\
  \enApp = \Package(\cont)
}{
  \jstepins[\enCiRet{\val}{\thunk}]{ \enState[ \val ][ \env ][ \store ][ \eventmap ][ \callinmap ][ \enFrame{\thunk}{\cont} ] }%
        { \enState[ \val ][ \env ][ \store ][ \eventmap ][ \callinmap ][ \cont ] }%
}
\;
\inferrule[ReturnCallback]{
\enThunk{\enClosure{ (\enFun[\enApp]{\var'}{\expr'}) }{\env'}}{\val'} = \thunk
  \\
  \enFwk = \Package(\cont)
}{
  \jstepins[\enCbRet{\val}{\thunk}]{ \enState[ \val ][ \env ][ \store ][ \eventmap ][ \callinmap ][ \enFrame{\thunk}{\cont} ] }%
        { \enState[ \val ][ \env ][ \store ][ \eventmap ][ \callinmap ][ \cont ] }%
}
\;
\inferrule[InvokeDisallowed]{
  \thunk \in \callinmap
}{
  \jstepins[\enDis{\enCi{\thunk}}]%
        { \enState[ \enInvoke{\thunk} ] }%
        \enDisallowed
}
\\
\thunkof(\thunk) \defeq \thunkof(\enFrame{\thunk}{\cont}) \defeq \thunk
\and
\thunkof(\enLetK{\var}{\cont}{\expr}{\env}) \defeq \thunkof(\cont)
\and
 \Package(\cont) \defeq \text{$\pkg$ if $\enThunk{\enClosure{ (\enFun[\pkg]{\var}{\expr}) }{\env}}{\val} = \thunkof(\cont)$}
\end{mathpar}
\caption{The instrumented transition relation
  $\jstepins[\glabel{}]{\stateconc}{\stateconc'}$ defines the app-framework interface and observing the event-driven protocol.}
\label{fig:model-semantics-instrumented}
\end{figure}

In \figref{model-semantics-instrumented}, we define the judgment form
$\smash{\jstepins{\stateconc}{\stateconc'}}$, which instruments our
small-step transition relation $\jstep{\stateconc}{\stateconc'}$ with
message $\glabel{}$.
%
%
Recall from \secref{overview} that we define a callback as an
invocation that transitions from framework to app code and a callin as
an invocation from app to framework code.
In $\semname$, this definition is captured crisply by the execution context $\cont$ in
which a thunk is forced.
In particular, we say that a thunk $\thunk$ is a callback invocation
$\enCb{\thunk}$ if the underlying callee function is an app function
(package \enkwApp{}), and it is called from a framework function
(package \enkwFwk{}) as in rule \TirName{ForceCallback}.
The $\thunkof(\cdot)$ function inspects the continuation for the running,
caller thunk. The $\Package(\cdot)$ function gets the package of the
running thunk.

Analogously, a thunk $\thunk$ is a callin
$\enCi{\thunk}$ if the callee function is in the \enkwFwk{} package, and
the caller thunk is in the \enkwApp{} package via rule \TirName{ForceCallin}.
%
\begin{example}[Observing a Callback] Letting \codej{handlePostExecute} be a framework function (i.e., in package $\enkwFwk$) and
\codej{onPostExecute} be an $\enkwApp$ function, the observable transition from the framework to the app defines the forcing of \codej{cb} as a callback:
\par\noindent\begin{minipage}{\linewidth}\centering
\usebox{\SBoxEnPostExecute}
\end{minipage}
\end{example}

In the above, we focused on the transition back-and-forth between framework and app code via calls. Returning from calls can also be seen as a ``message exchange'' with a return from a callin as another kind of \emph{back-message} going from framework code to app code (left-to-right in the figures from \secref{overview}). We write a callin-return back-message
$\enCiRet{\val}{\thunk}$ indicating the returning thunk $\thunk$ with return value $\val$. Likewise, a return from a callback is another kind \emph{in-message} going from app code to framework code (right-to-left).
We instrument returns in a similar way to forcings with the return back-message with \TirName{ReturnCallin} and the return in-message with \TirName{ReturnCallback}.
%

Finally to make explicit protocol violations, we instrument the \TirName{InvokeDisallowed} rule to record
the disallowed-callin invocations.
%
These rules replace the corresponding rules \TirName{Force},
\TirName{Return}, and \TirName{InvokeDisallowed} from
\figref{model-semantics-event}.
For replacing the \TirName{Force} and
\TirName{Return} rules, we elide two
rules, one for each, where there is no switch in packages
(i.e., $\pkg' = \Package(\cont)$ where $\pkg'$ is the package of the callee message).
%
%
%
%
These ``uninteresting'' rules and the remaining rules defining
the original transition relation $\jstep{\stateconc}{\stateconc'}$ not
discussed here are simply copied over with an empty message label
$\emptyword$.
\subparagraph{Observable Traces and Dynamic-Analysis Instrumentation.}

As described above, the app-framework interface is defined by the
possible messages that can exchanged where messages consist of
callback-callin invocations and their returns. A possible
app-framework interaction is thus a trace of such observable messages.

\begin{definition}[App-Framework Interactions as Observable Traces]
Let $\denote{\expr}$ be the path semantics of \semname{}
expressions $\expr$ that collects the finite sequences
of alternating state-transition-state $\stateconc\glabel{}\stateconc'$
triples according to the instrumented transition relation
$\smash{\jstepins[\glabel{}]{\stateconc}{\stateconc'}}$.
Then, an \emph{observable trace} is a finite sequence of messages
$\mword{}\colon \glabel1 \ldots \glabel n$ obtained
from a path by dropping the intermediate states and keeping the non-$\emptyword$ messages.
We write $\traceset{\expr}$ for the set of the observable traces
obtained from the set of paths, $\denote{\expr}$, of an expression $\expr$.
\end{definition}


An observable trace $\mword{}$ \emph{violates} the
event-driven application-programming protocol if $\mword{}$ ends with a disallowed
$\enkwDis$
message.



These definitions yield a design for a dynamic-analysis
instrumentation that observes app-framework interactions.
The trace recording in \toolname{} obtains observable traces
$\mword{}$ like the app-framework dialogue diagrams in \secref{overview}
by following the
instrumented semantics
$\smash{\jstepins[\glabel{}]{\stateconc}{\stateconc'}}$.
\toolname{} maintains a stack similar to the
continuation $\cont$ to emit the messages corresponding to the
forcings and returns of callbacks and callins,
and it emits disallowed $\enDis{\thunk}$ messages by observing the exceptions thrown by
the framework.


\section{Specifying Protocols and Modeling Callback Control Flow}
\label{sec:specification}

Using $\semname$ as a concrete semantic foundation, we first formalize an
abstraction of event-driven programs composed of separate app and framework
code with respect to what is observable at the app-framework
interface. This abstract transition system captures the possible
enabled-event and disallowed-call stores internal to the framework
that are consistent with observable traces, essentially defining a family of lifestate framework abstractions.
%
Then, we instantiate this definition for a specific lifestate language
that both specifies event-driven application-programming
protocols and models callback control flow.

The main point in these definitions is that lifestate modeling of callback control flow can only depend on what is observable at the app-framework interface.
Furthermore, the concrete
semantic foundation given by $\semname$ leads to a careful definition
of soundness and precision and a basis for model validation 
and
predictive-trace verification (\secref{verification}).


%
%
%

\begin{figure*}[tb]\footnotesize\centering
  \begin{minipage}{\linewidth}
  \small\begin{mathpar}
  %
  \text{states} \quad
  \absstate 
  \bnfdef \enAbsState \bnfalt \enDisallowedTrace
  \and
  \text{permitted-back} \quad\perset \bnfdef \cdot \bnfalt \perset\extMsg{\glabelFwk{}}
  \and
  \text{prohibited-in} \quad\proset \bnfdef \cdot \bnfalt \proset\extMsg{\glabelApp{}}
 \end{mathpar}
 \hfill\fbox{$\jstepnoins[\glabel{}]{\absstate}{\absstate'}$}
 \footnotesize\begin{mathpar}
  \inferrule[PermittedBack]{
    \begin{array}{@{}c@{}}
    \glabelFwk{} \in \perset \quad
    \mword{}' = \mword{} \glabelFwk{} \arcr
		\perset' = \evalFwk{\perset}{\rspec{}}{\mword{}'} \quad
		\proset' = \evalApp{\proset}{\rspec{}}{\mword{}'}
      \end{array}
  }{
    \jstepnoins[\glabelFwk{}]
          { \enAbsState }%
          { \enAbsState
            [\perset']
						[\proset']
            [\mword{}'] }
  }

	\inferrule[ProhibitedIn]{
    \begin{array}{@{}c@{}}
    \glabelApp{} \in \proset \arcr
    \mword{}' = \mword{} (\enDis{\glabelApp{}})
    \end{array}
  }{
    \jstepnoins[\enDis{\thunk}]
          { \enAbsState }%
          { \enDisallowedTrace[\mword{}'] }
  }

  \inferrule[PermittedIn]{
    \begin{array}{@{}c@{}}
    \glabelApp{} \notin \proset \quad
    \mword{}' = \mword{} \glabelApp{} \arcr
		\perset' = \evalFwk{\perset}{\rspec{}}{\mword{}'} \quad
		\proset' = \evalApp{\proset}{\rspec{}}{\mword{}'}
    \end{array}
  }{
    \jstepnoins[\glabelApp{}]
          { \enAbsState }%
          { \enAbsState
          [\perset']
					[\proset']
          [\mword{}']}
  }
  \end{mathpar}
  \end{minipage}
\caption{
  This transition system defines an abstraction of the
  framework-internal state consistent with an observable trace
  $\mword{}$ with respect to a framework abstraction $\rspec{}$.
  %
    %
    %
  The abstract state $\absstate$ contains a store of permitted
  back-messages $\perset$
  and a store of prohibited in-messages $\proset$, corresponding to
  an abstraction of enabled events and
  disallowed calls, respectively. The meaning of the framework abstraction
  $\rspec{}$ is captured by
    the store-update functions
    $\updatename{\enFwkSub}{\rspec{}}$ and $\updatename{\enAppSub}{\rspec{}}$,
    which determine how
    an abstract store changes on a new message.
  }
\label{fig:abs-message}
\end{figure*}

\subparagraph{Abstracting Framework-Internal State by Observing Messages}
\label{sec:abs-framework-internal}
\JEDI{We define an abstraction that tracks the set of messages that
  can be exchanged by the app and the framework.}
In~\figref{abs-message}, we define the transition system that abstracts the
framework-internal state consistent with an observable trace $\mword{}$.
%
%
%
An abstract state $\enAbsState$ contains
a store of \emph{permitted back-messages} $\perset$ and
a store of \emph{prohibited in-messages} $\proset$.
%
%
What the transition system captures are the possible traces consistent with iteratively applying a framework abstraction $\rspec{}$ to the current abstract state:
it performs a transition with a
back-message $\glabelFwk{}$ only if $\glabelFwk{}$ is permitted $\glabelFwk{} \in \perset$,
and a transition with an in-message $\glabelApp{}$
only if $\glabelApp{}$ is not prohibited $\glabelApp{} \not \in \proset$.
The trace $\mword{}$ in an abstract state saves the history of messages observed so far.
%
%
%
%
In the most general setting for modeling the event-driven framework,
the transition system can update the stores $\perset$ and  $\proset$ as
a function of the history of the observed messages $\mword{}$.
These updates are
formalized with the store-update functions
$\smash{\evalFwk{\perset}{\rspec{}}{\mword{}}}$ and $\smash{\evalApp{\proset}{\rspec{}}{\mword{}}}$
that define
an abstraction $\rspec{}$ of the event-driven framework describing
both its
application-programming protocol and its callback control flow.
%
%
%
A framework abstraction $\rspec{}$ also defines the initial abstract state
$\smash{\enAbsState[\persetinit][\prosetinit][\emptyword]}$ that contains
the initial (abstract) state of the stores of the permitted back-messages and prohibited in-messages.

%



The semantics $\traceset{\rspec{}}$ of a framework abstraction
$\rspec{}$ is the set of observable traces of the transition system
defined in \figref{abs-message} instantiated with $\rspec{}$. We get sequences of states from the transition relation
$\jstepnoins{\absstate}{\absstate'}$, read the observable trace
$\mword{}$ from the final state, and form a set of all such observable traces.

\begin{definition}[Soundness of a Framework Abstraction]
\begin{inparaenum}[(1)]
  \item A framework abstraction $\rspec{}$ is a sound abstraction of a $\semname$
program $\expr$ if $\traceset{\expr} \subseteq \traceset{\rspec{}}$;
  \item A framework abstraction $\rspec{}$ is a sound abstraction of a
framework implementation $\FwkImpl$ if and only if $\rspec{}$ is sound for every possible program
$\expr$ that uses the framework implementation $\FwkImpl$.
\end{inparaenum}
\end{definition}

The possible observable traces of a framework abstraction $\rspec{}$
is slightly richer than observable $\semname$ traces in that
callback-return messages $(\enCbRet{\val}{\thunk})$ may also be
prohibited (in addition callin-invocations
$\enCi{\thunk}$). Prohibiting callback-return messages corresponds to
specifying a protocol
where the app yields an invalid return value. If we desire to
capture such violations at the concrete level, it is straightforward
to extend $\semname$ with ``return-invalid'' transitions by analogy to
\TirName{InvokeDisallowed} transitions.
%
%

Finally, if $\rspec{1}$ and $\rspec{2}$ are sound specifications, we say
that $\rspec{1}$ is \emph{at least as precise} as $\rspec{2}$ if
$\traceset{\rspec{1}} \subseteq \traceset{\rspec{2}}$.

\begin{figure}
\subcaptionbox{%
  A lifestate abstraction is a set of rules that permits ($\renable$)
  or prohibits ($\rdisable$) parametrized messages $\mlabel{}$.
  %
	\label{fig:syntax-ls}
  }[0.98\linewidth]{
  \begin{minipage}{\linewidth}

  %
  \footnotesize\begin{mathpar} 
  \text{parametrized messages}\quad \mlabel{} \bnfdef
  \cbentry{\fun}{\mparams}
  \bnfalt \ciexit{\fun}{\mparams}{\mparams'}
  \bnfalt \cientry{\fun}{\mparams}
  \bnfalt \cbexit{\fun}{\mparams}{\mparams'}
  \end{mathpar}
  \begin{mathpar}
  \begin{grammar}[@{}l][]
  lifestate rules &  \rrule{} &\bnfdef &\rregexp{} \renable \mlabel{}
  \bnfalt \rregexp{} \rdisable \mlabel{}
  \\
  lifestate abstractions & \rspec{} & \bnfdef & \cdot \bnfalt \rrule{} \rspec{}
  \\
  trace matchers: regular expressions of parametrized messages & \rregexp{}
  %
  \end{grammar}
  \end{mathpar}
  \footnotesize\begin{mathpar} 
		 \inferrule
	  {\text{symbolic variables}\quad \specvar \in \SpecVarSet
	  \and
	  \text{parameters}\quad \mparams \in \SpecVarSet \cup \ValSet
	  \and
	  \text{binding maps}\quad \binding{} \bnfdef
	  \cdot \bnfalt  \binding{}\ext{\maplet{\specvar}{\val}}}{}
  \end{mathpar}
  \end{minipage}
}
\par\subcaptionbox{%
Semantics of a lifestate framework abstraction. The store-update functions
  $\updatename{\enFwkSub}{\rspec{}}$ and
  $\updatename{\enAppSub}{\rspec{}}$
  find rules from $\rspec{}$ that match the given trace $\mword{}$
  and update the store according
  to a consistent binding $\binding{}$ from symbolic variables to values.
	\label{fig:spec-update}
}[0.98\linewidth]{
\begin{minipage}{\linewidth}\footnotesize\smallskip
\smallskip\par\noindent\begin{mathpar}
\inferrule
{\arraycolsep=1.1pt\def\arraystretch{1}
\begin{array}{@{}rcl@{}}
\evalFwk{\perset}{\rspec{}}{{\mword{}}} &
\defeq
&
\SetST{
\glabelFwk{}
}{
\consistent{\rspec{}}{\mword{}} \land
\left(
\neg \prohibit{\rspec{}}{\mword{}}{\glabelFwk{}} \land
(\permit{\rspec{}}{\mword{}}{\glabelFwk{}} \lor \glabelFwk{} \in \perset)
\right)
} \arcr 
\evalApp{\proset}{\rspec{}}{{\mword{}}} &
\defeq &
\SetST{
\glabelApp{}
}{
\consistent{\rspec{}}{\mword{}} \rightarrow
\left(
\neg \permit{\rspec{}}{\mword{}}{\glabelApp{}} \land
(\prohibit{\rspec{}}{\mword{}}{\glabelApp{}} \lor \glabelApp{} \in \proset)
\right)
}
\arcr [1ex] 
\permit{\rspec{}}{\mword{}}{\glabel{}} & \defeq &
\exists \rregexp{} \renable \mlabel{} \in \rspec{},
\exists \binding{},
(\mword{},\binding{} \models \rregexp{})
\land \applybind{\binding{}}{\mlabel{}} = \glabel{}\arcr
\prohibit{\rspec{}}{\mword{}}{\glabel{}} & \defeq &
\exists \rregexp{} \rdisable \mlabel{} \in \rspec{},
\exists \binding{},
(\mword{},\binding{} \models \rregexp{})
\land \applybind{\binding{}}{\mlabel{}} = \glabel{}\arcr
\consistent{\rspec{}}{\mword{}} & \defeq &
\forall \glabel{} \in \GMsgSet,
(\permit{\rspec{}}{\mword{}}{\glabel{}} \leftrightarrow
\neg \prohibit{\rspec{}}{\mword{}}{\glabel{}}) \arcr [1ex] 
\end{array}}{}
\end{mathpar}
\smallskip\par\noindent\begin{mathpar}
  \persetinit \;\defeq\; \evalFwk{\GMsgSetFwk}{\rspec{}}{\emptyword}
  \and
  \prosetinit \;\defeq\; \evalApp{\emptyset}{\rspec{}}{\emptyword}
\end{mathpar}
\end{minipage}
}
\caption{Lifestate is a
  language for simultaneously specifying event-driven protocols and modeling callback control
  flow in terms of the observable app-framework interface.}
\label{fig:lifestate-syntax-semantics}
\end{figure}

\subparagraph{A Lifestate Abstraction.}
\label{sec:lifestate-abstractions}

We arrive at lifestates by instantiating the framework abstraction
$\rspec{}$ in a direct way as shown in \figref{lifestate-syntax-semantics}.
%
%
%
To describe rules independent of particular programs or executions,
we \emph{parametrize messages} with symbolic variables
$\specvar \in \SpecVarSet$.
%
The definition of the parametrized messages $\mlabel{}$
is parallel to the non-parameterized version but
using parameters instead of simply concrete values $\val$.
%
%
We call a message $\mlabel{}$ \emph{ground} when it
does not have symbolic variables (from $\SpecVarSet$), and
we distinguish the ground and parameterized messages by using
normal $\glabel{}$ and bold $\mlabel{}$ fonts, respectively.
For example, the parametrized callback-invocation
message~$\cbentry{\fun}{\specvar}$
specifies that a callback function $\fun$
is invoked with an arbitrary value from $\ValSet$.
The variable $\specvar$ can be used across several messages in a rule,
expressing that multiple messages are invoked with, or return, the
same value.

A lifestate abstraction $\rspec{}$ is a set of rules, and a
\emph{rule} consists of trace matcher
$\rregexp{}$
that when matched either \emph{permits} ($\renable$ operator) or
\emph{prohibits} ($\rdisable$ operator) a
parametrized
message
$\mlabel{}$.
As just one possible choice for the matcher $\rregexp{}$, we consider
$\rregexp{}$ to be a regular expression where the symbols of
the alphabet are parametrized messages $\mlabel{}$.
%
%
%
In matching a trace $\mword{}$ to a regular expression of parametrized
messages, we obtain a \emph{binding} $\binding{}$ that maps symbolic
variables from the parametrized messages to the concrete values from
the trace.
%
%
Given a binding $\binding{}$ and a message
$\mlabel{}$, we write
$\applybind{\binding{}}{\mlabel{}}$ to denote the
message $\mlabel{}'$ obtained by replacing each symbolic variable $\specvar$
in $\mlabel{}$ with $\binding{}(\specvar)$ if defined.
%


The semantics of lifestates is given by a choice of store-update
functions $\updatename{\enFwkSub}{\rspec{}}$ and
$\updatename{\enAppSub}{\rspec{}}$ in \figref{spec-update}
and the abstract transition relation $\jstepnoins{\absstate}{\absstate'}$
defined previously in \secref{abs-framework-internal}. The
store-update functions work intuitively by matching the given trace $\mword{}$
against the matchers $\rregexp{}$ amongst the rules in $\rspec{}$ and
then updating the store according to the matching rules
$\Set{\rrule{1}, \ldots, \rrule{n}} \subseteq \rspec{}$.
%
%

To describe the store-update functions in \figref{spec-update},
we write $\mword{},\binding{} \models \rregexp{}$ to express that a
trace $\mword{}$ and a binding $\binding{}$ satisfy a regular
expression $\rregexp{}$.
The definition of this semantic relation is standard, except for parametrized
messages $\mlabel{}$.
Here, we explain this interesting case
for when the trace $\mword{}$ and the binding $\binding{}$ satisfy the regular
expression $\mlabel{}$
(i.e., $\mword{},\binding{} \models \mlabel{}$):
\[\begin{array}{rcl}
  \mword{},\binding{} \models \mlabel{} & \text{iff} &
  \text{$\mword{} = \glabel{}$ and $\applybind{\binding{}}{\mlabel{}}
                                                       = \glabel{}$
                                                       for some ground
                                                       message $\glabel{}$}
\end{array}\]
A necessary condition for
$\mword{},\binding{} \models \mlabel{}$ is, for example,
that $\binding{}$ must assign a value to all the variables in
$\mlabel{}$, to get a ground message, and the message must be equal to
the trace $\mword{}$.
Note that, if there is no such ground message for $\mlabel{}$ with the
binding $\binding{}$, then $\mword{},\binding{} \not\models \mlabel{}$.
The full semantics of matching parametrized regular expressions is given in \refappendix{sec:appendix:regexp}.

Now, the function $\evalFwk{\perset}{\rspec{}}{{\mword{}}}$ captures how
the state of the permitted back-messages store $\perset$ changes according to the
rules $\rspec{}$.
As a somewhat technical point, a back-message can only be permitted if the rules $\rspec{}$ are \emph{consistent} with respect to the given trace $\mword{}$ (i.e.,
$\consistent{\rspec{}}{\mword{}}$).
The $\consistent{\rspec{}}{\mword{}}$ predicate holds iff there are no rules that
permits and prohibits $\glabel{}$ for the same message $\glabel{}$ and trace $\mword{}$.
Then, if the predicate $\consistent{\rspec{}}{\mword{}}$ is true,
the back-message $\glabelFwk{}$ must not be prohibited given the trace $\mword{}$
(i.e., $\neg \prohibit{\rspec{}}{\mword{}}{\glabelFwk{}}$).
Finally, if back-message $\glabelFwk{}$ is not prohibited, either it is permitted
by a specification for this trace $\mword{}$ (i.e., $\permit{\rspec{}}{\mword{}}{\glabelFwk{}}$)
or it was already permitted in the current store $\perset$ (i.e., $\glabelFwk{} \in \perset$).
%
%
The function $\evalApp{\proset}{\rspec{}}{{\mword{}}}$ is similar, but
it is defined for the prohibited in-messages store $\proset$. An in-message
$\glabelApp{}$ is prohibited first if the rules are not consistent.
Then, if the rules are consistent, the in-message $\glabelApp{}$ must
not be permitted by this trace, and either it is prohibited by a rule for this trace or the in-message was already prohibited in the current store $\proset$.
%
%
The auxiliary
predicates
$\permit{\rspec{}}{\mword{}}{\glabel{}}$
and $\prohibit{\rspec{}}{\mword{}}{\glabel{}}$
formally capture these conditions.
The $\permit{\rspec{}}{\mword{}}{\glabel{}}$ predicate is true iff
there is a rule $\rregexp{} \renable \mlabel{}$ in the specification
$\rspec{}$ that permits a message $\mlabel{}$
and a binding $\binding{}$,
such that the trace and the binding satisfy the regular expression
($\mword{},\binding{} \models \rregexp{}$),
and the ground message permitted
by the rule $\applybind{\binding{}}{\mlabel{}}$ is $\glabel{}$.
The $\prohibit{\rspec{}}{\mword{}}{\glabel{}}$ predicate is analogous but for
prohibit rules.

A key point is that the store-update functions
$\evalFwk{\perset}{\rspec{}}{{\mword{}}}$ and
$\evalApp{\proset}{\rspec{}}{{\mword{}}}$
are defined only in terms of what is observable at the app-framework interface $\mword{}$ and stores of permitted back-messages $\perset$ and prohibited in-messages $\proset$. Lifestate abstractions $\rspec{}$ do not depend on framework or app expressions $\expr$, nor framework-internal state.


\section{Dynamic Reasoning with Lifestates}
\label{sec:dynamic}

Lifestates are precise and detailed abstractions of
event-driven frameworks that
simultaneously specify the
protocol that the app
should observe and the callback control-flow assumptions that an app
can assume about the framework.
The formal development of lifestates in the above offers a clear approach for
\emph{model validation} and \emph{predictive-trace verification}.
In this section, we define the model validation and verification
problem and provide an intuition of their algorithms using the formal development in the previous sections. For
completeness of presentation, we provide further details in
\refappendix{sec:appendix:dynamic}.

\subparagraph{Validating Lifestate Specifications.}
\label{sec:formal-validation}
\JEDI{We cannot automatically check that a specification is sound, and
  hence we rely on simulation. Simulation allow us to trust the
  specification and to understand when it is wrong}%
As documentation in a real framework implementation like Android is
incomplete and ambiguous, it is critical that framework
abstractions have a mechanism to validate candidate rules---in a
manner independent of, say, a downstream static or dynamic analysis.

%
We say that a specification $\rspec{}$ is \emph{valid} for an
observable trace $\mword{}$ if $\mword{} \in \traceset{\rspec{}}$.
If a specification $\rspec{}$ is not valid for a trace $\mword{}$ from
a program $\expr$, then $\rspec{}$ is not a sound abstraction of
$\expr$.
%
%

We can then describe an algorithm that checks if $\rspec{}$ is a valid
specification for a trace $\mword{}$ with a reduction to a model
checking problem.
Lifestate rules specify the behavior
of an unbounded number of objects through the use of symbolic variables
$\specvar \in \SpecVarSet$ that are implicitly universally quantified
in the language and hence describe an unbounded number of messages.
However, as an observable trace $\mword{}$ has a finite number of
ground messages, the set of messages that we can use to instantiate the
quantifiers is also finite. Thus, the validation algorithm first ``removes''
the universal quantifier with the \emph{grounding} process that
transforms the lifestate abstraction $\rspec{}$ to a \emph{ground
abstraction} $\grspec{}$ containing only ground rules.

The language $\traceset{\grspec{}}$ of a ground specification
$\grspec{}$ can be represented with a finite transition system since
the set of messages in $\grspec{}$ is finite, and lifestate
rules are defined using regular expressions.
We then pose the validation problem as a model checking problem that we
solve using off-the-shelf symbolic model checking tools~\cite{DBLP:conf/cav/CavadaCDGMMMRT14}.
The transition system that we check
is the parallel composition (i.e., the intersection of the languages
of transition systems) of the transition system that accepts only
the trace $\mword{}$ and the transition system
$\jstepnoins{\absstate}{\absstate'}$
parametrized by the grounded lifestate abstraction $\grspec{}$.
The lifestate abstraction $\grspec{}$ is valid if and only if the composed
transition system reaches the last state of the trace $\mword{}$.

\subparagraph{Dynamic Lifestate Verification}
\label{sec:verification}

\JEDI{We define the dynamic verification problem for an execution
  trace of an event-driven program and how we can solve it by using
  the message abstraction}%
%
Because of the previous sections building up to
lifestate validation, the formulation of the dynamic verification
is relatively straightforward and offers a means to
evaluate the expressiveness of lifestate specification.
%

\JEDI{We split the trace to identify the list of single executions of
  the first-level callback executed, which correspond to event
  executions}
We define the set of sub-traces of a trace
$\mword{}=\rtracecb{1} \ldots \rtracecb{l}$
as
$\subtraceset{\mword{}} \defeq \{\rtracecb{1}, \ldots, \rtracecb{l}\}$,
where $\rtracecb{}'\in \subtraceset{\mword{}}$ if $\rtracecb{}'$ is a
substring of $\mword{}$ that represents the entire execution of a
callback directly invoked by an event handler.
%
%
%
We consider the set $\tracerepetitions$
of all the traces obtained by repeating the elements in
$\subtraceset{\mword{}}$ zero-or-more times and
%
%
\(
\repetitions{\mword{}}{\expr} \defeq
\traceset{\expr} \cap \tracerepetitions
\) its intersection with the traces of the \semname program
$\expr$.

\JEDI{dynamic verification problem: no arbitrary repetition of callbacks
  in the observable trace violates the protocol}
Given an observable trace $\mword{}$ of the program $\expr$
(i.e., $\mword{} \in \traceset{\expr}$), the \emph{dynamic verification problem} consists of proving
the absence of a trace $\mword{}' \in \repetitions{\mword{}}{\expr}$ that
violates the application-programming protocol.
%
Since we cannot know the set of traces $\traceset{\expr}$ for a
\semname program $\expr$ (i.e., the set of traces for the app composed with the framework implementation), we cannot solve the dynamic
verification problem directly. Instead, we solve an abstract version of
the problem, where we use a lifestate specification $\rspec{}$ to
abstract the framework implementation $\FwkImpl$.
Let
\(
\repetitions{\mword{}}{\rspec{}} \defeq \traceset{\rspec{}} \cap \tracerepetitions
\)
be the set of repetitions of the trace $\mword{}$ that can be
seen in the app-framework interface abstraction defined by $\rspec{}$.

\JEDI{The approach is sound: if we do not find any protocol violation
  in the abstract, then there are no protocol violation in the
  concrete}
Given a trace $\mword{} \in \traceset{\expr}$ and a sound specification
$\rspec{}$, the \emph{abstract dynamic verification problem}
consists of proving the absence of a trace $\mword{}' \in
\repetitions{\mword{}}{\rspec{}}$ that violates the
application-programming protocol.
%
If we do not find any protocol violation using a specification
$\rspec{}$,
then there are no
violations in the possible repetitions of the concrete trace
$\mword{}$.
Observe that the key verification challenge is getting a precise enough framework abstraction $\rspec{}$ that sufficiently restricts the possible repetitions of the concrete trace $\mword{}$.

We reduce the abstract dynamic verification problem to a model
checking problem in a similar way to validation:
we first generate the \emph{ground model} $\grspec{}$ from the
lifestate model $\rspec{}$ and the trace $\mword{}$. Then, we construct
the transition system that only generates traces in the set
$\repetitions{\mword{}}{\rspec{}}$ by composing the transition system
obtained from the ground specification $\grspec{}$ and the automaton
accepting words in $\tracerepetitions$.
This transition system satisfies a safety property iff
there is no trace $\mword{} \in \repetitions{\mword{}}{\rspec{}}$ that
violates the protocol.






\section{Empirical Evaluation}
\label{sec:expeval}

\newcommand{\rqone}{\emph{RQ1}\xspace}
\newcommand{\rqtwo}{\emph{RQ2}\xspace}
\newcommand{\rqthree}{\emph{RQ3}\xspace}
\newcommand{\rqfour}{\emph{RQ4}\xspace}
\newcommand{\expone}[0]{\emph{EXP1}\xspace}
\newcommand{\exptwo}[0]{\emph{EXP2}\xspace}
\newcommand{\justdis}{top}
\newcommand{\tflow}{flow}
\newcommand{\lifecycle}[0]{lifecycle}
\newcommand{\lifestatena}[0]{\emph{lifestate}\xspace}
\newcommand{\lifestateat}[0]{lifestate}
\newcommand{\baseline}{baseline}
\newcommand{\flowdroidmodel}{FlowDroid\xspace}
\newcommand{\verifiable}{verifiable}
\newcommand{\tabularh}[2]{\multicolumn{#1}{c}{#2}}
\newcommand{\nobind}{none}
\newcommand{\fixedbind}{fixed}
\newcommand{\fluidbind}{fluid}
\newcommand{\fulllifestate}{full}
\newcommand{\verified}{verified}

\newcommand{\tracecount}{$2202$\xspace}
\newcommand{\appcount}{$121$\xspace}

\newcommand{\noerror}{No errors\xspace}

\newcommand{\rmlifecyclebinding}[1]{}

\JEDI{We implement our ideas above in a real tool.}
We implement our approach for Android in the \toolname{}
tool that
\begin{inparaenum}[(i)]%
	\item instruments an Android app to record observable traces,
	\item validates a lifestate model for soundness against a corpus of traces, and
	\item assesses the precision of a lifestate model with dynamic verification.
\end{inparaenum}
\JEDI{Main challenge for edpp verification is framework modeling.}%
We use the following research questions to demonstrate that lifestate is an effective language to model event-driven protocols, and validation is a crucial step to avoid unsoundness.

%
%
\begin{enumerate}[RQ1]
  \item{\emph{Lifestate Precision.}} Is the lifestate language adequate to model the callback control flow of Android?
	The paper hypothesizes that carefully capturing the app-framework interface is necessary
	to obtain precise protocol verification results.
  \item{\emph{Lifestate Generality.}}
	Do lifestate models generalize across apps?
  We want to see if a lifestate model is still precise when used on a
  trace from a new, previously unseen app.

\item{\emph{Model Validation.}}
Is validation of callback control-flow models with concrete traces necessary to develop \emph{sound} models?
We expect to witness unsoundnesses in existing (and not validated) callback control-flow models and that validation is a crucial tool to get sound models.
%
\end{enumerate}

\noindent
Additionally, we considered the feasibility of continuous model validation. The bottom line is that we could validate 96\% of the traces within a 6 minute time budget; we discuss these results further in \refappendix{sec:appendix:exp:feasibility}.


\newcommand{\expsetfig}[1]{\multicolumn{10}{@{}c@{}}{#1}}
\newcommand{\rulesetfig}{\multicolumn{2}{@{}c@{}}{\flowdroidmodel} & & \multicolumn{2}{@{}c@{}}{\justdis} & & \multicolumn{2}{@{}c@{}}{\lifecycle} & & \multicolumn{2}{@{}c@{}}{\baseline} & &  \multicolumn{2}{@{}c@{}}{\lifestateat}}
\newcommand{\invci}{sensitive}
\newcommand{\resheadfig}{\multicolumn{2}{@{}c@{}}{false alarms}}
\newcommand{\senscallinfig}{callin}
\newcommand{\countunit}{(n)}
\newcommand{\percunit}{(\%)}
\newcommand{\ulfig}{\countunit & \countunit & & \countunit & \percunit & & \countunit & \percunit & & \countunit & \percunit}

\newcommand{\tblcallin}[1]{\scalebox{0.8}{#1}}

\begin{table}[b]\scriptsize
\caption{%
Precision of callback control-flow models.
The \emph{sensitive callin} column lists protocol properties by the callin that crashes the app when invoked in a bad state.
We collect a total of 50 traces from 5 applications with no crashes.
The \emph{sensitive} column lists the number of traces where the application invokes a sensitive callin.
To provide a baseline for the precision of a model, we count the number of traces without a manually-confirmed real bug in the \emph{verifiable} column.
There are four columns labeled \emph{verified} showing the number and percentage of verifiable traces proved correct using different callback control-flow models.
The \emph{lifestate} columns capture our contribution.
The \emph{\othermodel} columns capture the current practice for modeling the Android framework.
The \emph{bad} column lists the number of missed buggy traces and is discussed further in \rqtwo{}.%
}
\label{tbl:small_experiment}
\begin{tabular*}{\linewidth}{@{\extracolsep{\fill}}lrr*{3}{rr}rrr@{}}\toprule
properties
  & \tabularh{2}{non-crashing traces}
  & \tabularh{9}{callback control-flow models}
\\
\cmidrule{1-1}
\cmidrule{2-3}
\cmidrule{4-12}
  &
	&
  & \tabularh{2}{\justdis} 
  & \tabularh{2}{\lifecycle}
    & \tabularh{2}{\lifestateat}
  & \tabularh{3}{\othermodel{}}
\\
\cmidrule{4-5}
\cmidrule{6-7}
\cmidrule{8-9}
\cmidrule{10-12}
sensitive
  & \tabularh{1}{\invci} 
  & \tabularh{1}{\verifiable}
  & \tabularh{2}{\verified}
  & \tabularh{2}{\verified}
  & \tabularh{2}{\verified}
  & \tabularh{2}{\verified}
  & \tabularh{1}{bad}
\\
\senscallinfig
  & \countunit
  & \countunit
  & \countunit
  & \percunit
  & \countunit
  & \percunit
  & \countunit
  & \percunit
  & \countunit
  & \percunit
  & \countunit
\\
\cmidrule{1-1}
\cmidrule{2-3}
\cmidrule{4-12}


%
    \codej{AlertDialog}\\
    \tblcallin{dismiss}       & 16        & 6          & 0 &  0 & 0 &  0  & 6 &100 & 6 & 100 & 0\\
    \tblcallin{show}          & 43        &34          &17 & 50 &17 & 50  &28 & 82 & 24& 71 & 0\\
    \codej{AsyncTask}\\
    \tblcallin{execute}       &  4        & 4          & 0 &  0 & 4 &100  & 4 &100 & 0 & 0 & 0\\
    \codej{Fragment}\\
    \tblcallin{getResources}  & 10        &10          & 0 &  0 & 0 &  0  &10 &100 & 4 & 40 & 0\\
    \tblcallin{getString}     & 10        &10          & 0 &  0 & 0 &  0  & 2 & 20 & 0 & 0 & 0\\
    \tblcallin{setArguments}  & 19        &19          & 1 &  5 & 1 &  5  &19 &100 & 13& 68 & 0\\
\cmidrule{1-1}
\cmidrule{2-3}
\cmidrule{4-12}
    total                  &102        &83          &18 & 22 &22 & 27  &69 & 83 & 47 & 57 & 0 \\
\bottomrule
\end{tabular*}
\end{table}

\JEDI{We evaluate whether the precision of lifestate is needed for improvement over lifecycle.}

\subparagraph{\rqone{}: Lifestate Precision.}
\label{sec:rq1}


\JEDI{lifestate was able to reach the highest percentage of proven traces, flowdroid is a reasonable second, but may not be the best model as we may see later.}
The bottom line of \tblref{small_experiment} is that lifestate modeling is essential to improve the percentage of verified traces to 83\%---compared to 57\% for \othermodel{} and 27\% for lifecycle modeling.

%


\emph{Methodology.}
We collect execution traces from Android apps and compare the
precision obtained verifying protocol violations with
four different callback control-flow models.
%
The first three models are expressed using different subsets of the lifestate language.
The \emph{top} model is
the least precise (but clearly sound) model where any callback can happen at all times, like in the
Automaton~\ref{fig:feedremover-automata-top}
in \secref{overview}.
The \emph{lifecycle} model represents the most precise callback control-flow model that we can express only using back-messages, like in
Automaton~\ref{fig:feedremover-automata-lifecycle-noregistration}.
The \emph{lifestate} model uses the full lifestate language, and hence
also in-messages like in the
Automaton~\ref{fig:feedremover-automata-lifestate}, to change the
currently permitted back-messages.
It represents the most precise model that we can represent with lifestate.
To faithfully compare the precision of the formalisms, we improved the
precision of the \lifecycle{} and \lifestateat{} models minimizing the false
alarms from verification. And at the same time, we continuously run model validation to
avoid unsoundnesses, as we discuss below in \rqthree.
As a result of this process, we modeled the behavior of several
commonly-used Android classes, including \codej{Activity},
\codej{Fragment}, \codej{AsyncTask}, \codej{CountdownTimer},
\codej{View}, \codej{PopupMenu}, \codej{ListView}, and \codej{Toolbar}
and their subclasses.
Excluding similar rules for subclasses, this process resulted in a
total of 167 \lifestateat{} rules.

We further compare with an instance of a \emph{\othermodel}
model, which refines component lifecycles with callbacks from other Android objects.
%
Our model is a re-implementation of the model used in
FlowDroid~\cite{DBLP:conf/pldi/ArztRFBBKTOM14} that
considers the lifecycle for the UI components (i.e.,
\codej{Activity} and \codej{Fragment}) and
bounds the execution of a pre-defined list of callback methods
in the active state of the \codej{Activity} lifecycle, similarly to
the example we show in~\figref{activitylifecycle}.
We made a best effort attempt to faithfully replicate the \flowdroidmodel{} model (and discuss how we did so
in \refappendix{sec:appendix:exp:lifecyclepp}).

%
%

To find error-prone protocols, we selected \emph{sensitive
  callins}, shown in the first column of
\tblref{small_experiment}, that
frequently occur as
issues on GitHub and StackOverflow~\cite{antennapodbug,redreaderbug,android-topekabug,onebusawaybug,stackoverflow_dialog_dismiss,stackoverflow_fragment_setArguments}.
We then specify the lifestate rules to allow and disallow the sensitive callins.

To create a realistic trace corpus for \rqone,
we selected five apps by consulting Android user groups to find those that extensively use Android UI objects, are not overly simple (e.g., student-developed or sample-projects apps), and use at least one of the sensitive callins.
To obtain realistic interaction traces, we recorded manual
interactions from a non-author user who had no prior knowledge of the internals of the app.
The user used each app 10 times for 5 minutes (on an x86 Android
emulator running Android 6.0)---obtaining a set of 50 interaction
traces.
With this trace-gathering process, we exercise a wide range of behaviors of Android UI objects that drives the callback control-flow modeling.

To evaluate the necessity and sufficiency of lifestate, we compare the
verified rates (the total number of verified traces over the
total number of verifiable traces) obtained using each callback
control-flow model.
We further measure the verification run time to evaluate the trade-off between the expressiveness of the models and the feasibility of verification.

\emph{Discussion.}
In \tblref{small_experiment}, we show the number of verified traces and the verified rates broken down by sensitive callins and different callback control-flow models---aggregated over all apps.
As stated earlier,
the precision improvement with lifestate is significant,
essential to get to 83\% verified.
We also notice that the \rmlifecyclebinding{no-binding }lifecycle model is only slightly more precise
than the trivial \justdis{} model (27\% versus 22\% verified rate).
Even with unsoundnesses discussed later,
\othermodel{} is still worse than the lifestate model, with 57\% of traces proven.

Lifestate is also expressive enough to prove most verifiable traces---making manual triage of the remaining alarms feasible.
We manually examined the 14 remaining alarms with the lifestate model, and we identified two sources of imprecision:
\begin{inparaenum}[(1)]
	\item an insufficient modeling of the attachment of UI components (e.g., is a \codej{View} in the \codej{View} tree attached to a particular \codej{Activity}?), resulting in 13 alarms;
	%
	%
	%
	\item a single detail on how Android options are set in the app's XML, resulting in 1 alarm.
\end{inparaenum}
The former is not fundamental to lifestates but a modeling tradeoff where deeper attachment modeling offers diminishing returns on the verified rate while increasing the complexity of the model and verification times. The latter is an orthogonal detail for handling Android's XML processing (that allows the framework to invoke callbacks via reflection).

\subparagraph{\rqtwo: Lifestate Generality.}
\label{sec:rq2}

\begin{table*}[t]\scriptsize
	\caption{%
The table shows the precision results for the 1577 non-crashing traces
that contained a sensitive callins from a total of \tracecount traces that we collected from
	\appcount distinct open source app repositories. We note that lifestate takes slightly longer than lifecycle; for this reason, lifestate performs slightly worse than lifecycle for execute.
  The bad column
  is 0 for models other than \othermodel because of continuous validation. Note that out of 64 total buggy traces, \othermodel missed 27 bugs (i.e., had a 42\% false-negative rate).
}


	\label{tbl:big_experiment}
\begin{tabular*}{\linewidth}{@{\extracolsep{\fill}}lrr*{3}{rr}rrr@{}}\toprule
	properties
	  & \tabularh{2}{non-crashing traces}
	  & \tabularh{9}{callback control-flow models}
	\\
	\cmidrule{1-1}
	\cmidrule{2-3}
	\cmidrule{4-12}

	  &
		&
	  & \tabularh{2}{\justdis} 
	  & \tabularh{2}{\lifecycle}
	  & \tabularh{2}{\lifestateat}
	  & \tabularh{3}{\othermodel}
	\\
	\cmidrule{4-5}
	\cmidrule{6-7}
	\cmidrule{8-9}
	\cmidrule{10-12}
        sensitive
	  & \tabularh{1}{\invci} 
	  & \tabularh{1}{\verifiable}
	  & \tabularh{2}{\verified}
	  & \tabularh{2}{\verified}
	  & \tabularh{2}{\verified}
	  & \tabularh{2}{\verified}
	  & \tabularh{1}{bad}
	\\
	\senscallinfig
	  & \countunit
	  & \countunit
	  & \countunit
	  & \percunit
	  & \countunit
	  & \percunit
	  & \countunit
	  & \percunit
	  & \countunit
	  & \percunit
	  & \countunit
	\\
	\cmidrule{1-1}
	\cmidrule{2-3}
	\cmidrule{4-12}
	\codej{AlertDialog}\\
	\tblcallin{dismiss}    &  94 &  59 & 54 &92 & 54 & 92 &  54 & 92 & 58 & 98 & 3\\
	\tblcallin{show}       & 145 & 144 &125 &87 &124 & 86 & 125 & 87 &127 & 88 & 0\\
	\codej{AsyncTask}\\
	\tblcallin{execute}      & 415 & 415 &  0 & 0 &415 &100 & 412 & 99 &262 & 63 & 0\\
	\codej{Fragment}\\
	\tblcallin{getResources}  & 156 & 155 & 89 &57 & 89 & 57 & 128 & 83 &116 & 75 & 0\\
	\tblcallin{getString}     & 220 & 193 &124 &64 &124 & 64 & 134 & 69 &131 & 68 & 24\\
	\tblcallin{setArguments}  & 456 & 456 & 59 &13 &108 & 24 & 437 & 96 &435 & 95 & 0\\
	\tblcallin{startActivity} &  91 &  91 &  0 & 0 &  0 &  0 &  12 & 13 & 19 & 21 & 0\\
\cmidrule{1-1}
\cmidrule{2-3}
\cmidrule{4-12}
	total                  &1577 &1513 &451 &30 &914 & 60 &1302 & 86 & 1148 & 76 & 27\\
\bottomrule
\end{tabular*}
\end{table*}

The bottom line of \tblref{big_experiment} is that the lifestate model developed for \rqone{} as-is generalizes to provide precise results (with a verified rate of 86\%) when used to verify traces from \appcount previously unseen apps.
%
This result provides evidence that lifestates capture general behaviors of the Android framework.
While the \lifecyclepp{} model verifies 76\% of traces, it also misses 27 out of 64 buggy traces (i.e., has a 42\% false-negative rate).


\emph{Methodology.}
\JEDI{We collect 148 new apps by searching for open source
  apps which use our sensitive callins.}
%
%
To get a larger corpus, we cloned \appcount distinct open source
apps repositories from GitHub that use at least one sensitive
callin (the count combines forks and clones).
Then, we generated execution traces using the
Android UI Exerciser Monkey~\cite{exercise-monkey} that
interacts with the app issuing random UI events (e.g., clicks, touches).
We attempted to automatically generate three traces for each app file obtained by building each app.

\JEDI{We prove 23\% more traces with the lifestate rules over the lifecycle.}
\emph{Discussion.}
From \tblref{big_experiment}, we see that the \lifestateat{} verified
rate of 86\% in this larger experiment is comparable with the verified
rate obtained in \rqone.
Moreover, \lifestateat{} still improves the verified rate with respect to \lifecycle{}, which goes from 60\% to 86\%, showing that the expressivity of lifestate is necessary.

Critically, the \othermodel{} model does not alarm on 42\% of the traces representing real defects.
That is, we saw unsoundnesses of the \othermodel{} model manifest in the protocol verification client.

The verified rate for the \lifecycle{} model is higher in this larger corpus (60\%) compared to the rate in \rqone{} (27\%), and the precision improvement from the \justdis{} abstraction is more substantial (60\% to 30\% versus 27\% to 22\%). This difference is perhaps to be expected when using automatically-generated traces that may have reduced coverage of app code and
bias towards shallower, ``less interesting'' callbacks associated with application initialization instead of user interaction.
%
In these traces, it is possible that UI elements were not exercised as frequently, which
would result in more traces provable solely with the \lifecycle{} specification. Since coverage
is a known issue for the Android UI Exerciser Monkey~\cite{DBLP:conf/apsec/ArnatovichNTS16}), it was critical to have some evidence on deep, manually-exercised traces as in \rqone{}.

\emph{Bug Triage.}
We further manually triage every remaining alarm from both \rqone and \rqtwo.
Finding protocol usage bugs was not necessarily expected: for \rqone, we selected seemingly well-developed, well-tested apps to challenge verification, and for \rqtwo, we did not expect automatically generated traces to get very deep into the app (and thus deep in any protocol).






Yet from the \rqone triage, we found 2 buggy apps out of 5 total. These apps were Puzzles~\cite{boyle-dismissProgress} and SwiftNotes~\cite{fdroid-swiftnotes}. Puzzles had two bugs, one related to \codej{AlertDialog.show} and one for \codej{AlertDialog.dismiss}. Swiftnotes has a defect related to \codej{AlertDialog.show}.

In the \rqtwo{} corpus, we found 7 distinct repositories with a buggy app (out of \appcount distinct repositories) from 64 buggy traces (out of \tracecount). We were able to reproduce bugs in 4 of the repositories and strongly suspect the other 3 to also be buggy.
Three of the buggy apps invoke a method on \codej{Fragment} that requires the \codej{Fragment} to be attached. This buggy invocation happens within unsafe callbacks.  Audiobug~\cite{audiobug} invokes \codej{getResources}. NextGisLogger~\cite{nextgislogger} and Kistenstapeln~\cite{kistenstapeln} invoke \codej{getString}. We are able to reproduce the Kistenstapeln bug.

Interestingly, one of the apps that contain a bug is Yamba~\cite{yamba-git}, a tutorial app from a book on learning Android~\cite{Gargenta:2014aa}. We note that the Yamba code appears as a part of three repositories where the code was copied (we only count these as one bug). The tutorial app calls \codej{AlertDialog.dismiss} when an
\codej{AsyncTask} is finishing and hence potentially after the
\codej{Activity} object used in the \codej{AlertDialog} is not visible anymore.
%
We found similar defects in several actively maintained open
source apps where callbacks
in an \codej{AsyncTask} object were used either to invoke
\codej{AlertDialog.show} or \codej{AlertDialog.dismiss}. These apps included OSM Tracker~\cite{osmtracker} and Noveldroid~\cite{noveldroid-bug}.  Additionally, we found this bug in a binary library connected with the PingPlusPlus android app \cite{PingPlusPlus}.
By examining the output of our verifier, we were able to create a test to concretely witness defects in 4 of these apps.





\begin{figure}[t]\centering
    \includegraphics[width=0.45\linewidth]{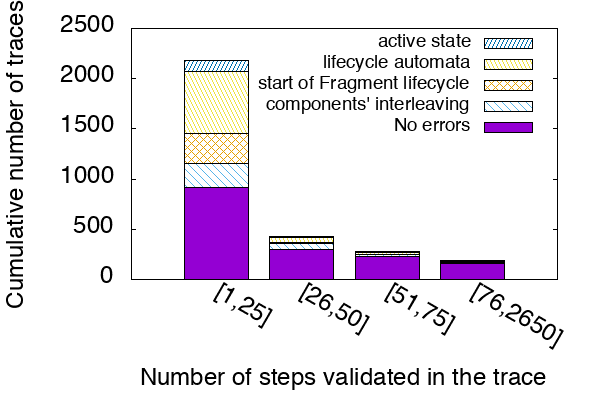}
\caption{Results of the validation of the \othermodel{} model on
  all the traces. We plot the cumulative traces grouped by (intervals of) the number of steps validated.
  The number of traces are further divided into categories, either indicating
  that validation succeeded, ``no errors,'' or the cause of failure
  of the validation process.
}
\label{fig:flowdroid-validation}
\end{figure}

\subparagraph{\rqthree: Model Validation.}
\label{sec:exp:validation}
\JEDI{Validation is neccessary because the FlowDroid model fails to validate on 58\% of traces.}
The plot in~\figref{flowdroid-validation} highlights the
necessity of applying model validation: \othermodel{} based on a widely used
callback control-flow model
does not validate (i.e., an unsoundness is witnessed) on 58\% of $2183$ traces (and the validation ran out of memory for $19$ out of the total $2202$ traces).
%
%

\emph{Methodology.}
%
We first evaluate the need for model validation by applying our
approach to \othermodel and quantifying its discrepancies with the
real Android executions.
%

%
Our first experiment validates the \othermodel{} model on all
the traces we collected (bounding each validation check to 1
hour and 4 GB of memory).
We quantify the necessity of model validation collecting for
each trace if the model was valid and the length of the
maximum prefix of the trace that the model validates.
%
Since there are already some known limitations in the \othermodel{}
model (e.g., components interleaving), we triage the results to
understand if the real cause of failure is a new mistake discovered
with the validation process.


Our second experiment qualitatively evaluates the necessity of model
validation to develop sound lifestate specifications.
To create a sound model, we started from the empty
model (without rules) and continuously applied validation to find
and correct mistakes. In each iteration:
we model the callback control flow for a specific Android object;
we validate the current model on the entire corpus of traces (limiting each trace to one hour and 4 GB of memory);
and when the model is not valid for a trace, we inspect the validation result and repair the specification.
We stop when the model is valid for all the traces.
We then collected the mistakes we found with automatic validation
while developing the \lifestateat{} model.
We describe such mistakes and discuss how we used validation to
discover and fix them.

\emph{Discussion: \othermodel Validation.}
%
%
%
%
From the first bar of the plot
in~\figref{flowdroid-validation}, we see that the
\othermodel{} model validates only 42\% of the total traces, while
validation fails in the remaining cases (58\%). The bar shows the number of
traces that we validated for at least one step, grouping them by
validation status and cause of validation failure.
From our manual triage, we identified $4$ different broad causes for
unsoundness:
\begin{inparaenum}[i)]
\item \emph{outside the active lifecycle}: the model
  prohibits the execution of a callback outside the
  modeled active state of the \codej{Activity};
\item \emph{wrong lifecycle automata}:
  the model wrongly prohibits the execution of an \codej{Activity} or
  \codej{Fragment} lifecycle callback;
\item \emph{wrong start of the \codej{Fragment} lifecycle}: the model
  prohibits the start of the execution of the \codej{Fragment}
  lifecycle;
\item \emph{no components interleaving}:
  the model prohibits the interleaved execution of callbacks from
  different \codej{Activity} or \codej{Fragment} objects.
\end{inparaenum}
%
The plot shows that the \othermodel{} model is not valid
on 25\% of the traces because it does not model the interleaving of
components (e.g., the execution of callbacks from different
\codej{Activity} and \codej{Fragment} objects cannot interleave%
) and the start of the
\codej{Fragment} lifecycle at an arbitrary point in the enclosing
\codej{Activity} object. With \flowdroidmodel{}, such limitations are known
and have been justified as practical choices to have
feasible flow analyses~\cite{DBLP:conf/pldi/ArztRFBBKTOM14}.
But the remaining traces, 33\% of the total, cannot be
validated due other reasons including modeling mistakes.
In particular, the \flowdroidmodel{}  model imprecisely captures
the lifecycle automata (for both \codej{Activity} and
\codej{Fragment}) and erroneously confines the execution of some
callbacks in the active state of the lifecycle.

The other bars in the plot of~\figref{flowdroid-validation}
show the number of traces we validated for more than 25, 50, and 75
steps, respectively.
In the plot, we report the total number of steps in the execution
traces that correspond to a callback or a callin that we either
used in the \lifestateat{} or the \othermodel{} model, while we remove
all the other messages.
From such bars, we see that we usually detect the
unsoundness of the \othermodel{} model ``early'' in the
trace (i.e., in the first 25 steps). This result is not surprising
since most of the modeling mistakes we found are related to the
interaction with the lifecycle automata
and can be witnessed in the first iteration of the lifecycle.
We further discovered that the
\othermodel{} model
mostly validates shorter execution traces,
showing that having sound models for
real execution traces is more challenging,
which we discuss further in \refappendix{flowdroid-validation-length}.
%

\emph{Discussion: Catching Mistakes During Modeling.}
We were able to obtain a valid lifestate specification \emph{for over 99.9\%} of the traces
in our corpus. That is, we were able to understand and model the objects we selected in all but two traces.

Surprisingly, we identified and fixed several mistakes in our
modeling of the \codej{Activity} and \codej{Fragment} lifecycle that
are due to undocumented Android behaviors.
An example of such behavior is the effect of \codej{Activity.finish} and \codej{Activity.startActivity} on the
callback control flow for the \codej{onClick} callback.
%
It is unsound to restrict
the enabling of \codej{onClick}
callbacks
to
the active state of the \codej{Activity} lifecycle (i.e.,
between the execution of the \codej{onResume} and \codej{onPause}
callbacks). This is the behavior represented with
blue edges in \figref{activitylifecycle},
what is typically understood from the Android documentation,
and captured in the existing
callback control-flow models used for static analysis.
%
\newcommand{\actv}{$\pa[a]$}
\newcommand{\activityattached}[2]{\textbf{activityAttached}(#1,#2)}
\newcommand{\listenerattached}[2]{\textbf{listenerAttached}(#1,#2)}

\newsavebox{\SBoxOnDestroyr}
\codejmk[\footnotesize]{\SBoxOnDestroyr}{(a:Activity).onDestroy()}
\newsavebox{\SBoxFindViewById}
\codejmk[\footnotesize]{\SBoxFindViewById}{(b:Button) = findViewById( i:int)}
\newsavebox{\SBoxFinish}
\codejmk[\footnotesize]{\SBoxFinish}{(a:Activity).finish()}

%
We implemented a model where \codej{onClick} could be invoked only when its \codej{Activity} was running and found this assumption to be invalid on several traces.
We inferred that the mistake was due to the wrong
``bounding'' of the \codej{onClick} callbacks in the \codej{Activity}
lifecycle since in all the traces:
\begin{inparaenum}[i)]
\item the first callback that was erroneously disabled in the model
  was the \codej{onClick} callback; and
\item the \codej{onClick} callback was disabled in the model just
  after the execution of an \codej{onPause} callback that appeared
  before in the trace, without an \codej{onResume} callback in between
  (and hence, outside the active state of the \codej{Activity}.)
\end{inparaenum}
%
%
%
It turns out that both \codej{finish} and \codej{startActivity} cause the \codej{Activity} to
pause without preventing the pending \codej{onClick} invocations from
happening, as represented in the \textcolor{red}{red} edges connected to \codej{onClick}
in~\figref{activitylifecycle}. We validated such behaviors by writing
and executing a test application and finding its description in
several Stack Overflow posts~\cite{click-after-pause,click-pause-2}.
%
The fix for this issue is to detect the
finishing state of the \codej{Activity} and to not disable the \codej{onClick} callback in this case.

%

\section{Related Work}
\label{sec:relwork}

\newcommand{\fmttool}[1]{#1}
\newcommand{\eventracer}{\fmttool{EventRacer}\xspace}
\newcommand{\emexplorer}{\fmttool{Em-Explorer}\xspace}
\newcommand{\rfour}{\fmttool{r4}\xspace}
\newcommand{\flowdroid}{\fmttool{FlowDroid}\xspace}
\newcommand{\hopper}{\fmttool{Hopper}\xspace}
\newcommand{\scandal}{\fmttool{Scandal}\xspace}
\newcommand{\symdroid}{SymDroid\xspace}
\newcommand{\edgeminer}{\fmttool{EdgeMiner}\xspace}
\newcommand{\lithium}{\fmttool{Lithium}\xspace}
\newcommand{\tracematch}{\fmttool{TraceMatch}\xspace}
\newcommand{\droidel}{\fmttool{Droidel}\xspace}
\newcommand{\pasket}{\fmttool{Pasket}\xspace}

Several works~\cite{DBLP:conf/pldi/ArztRFBBKTOM14,DBLP:conf/oopsla/BlackshearCS15,gator-toolkit,DBLP:conf/vstte/SmithC15,Jeon2012SymDroidSE,DBLP:conf/icse/RadhakrishnaLMM18,DBLP:conf/icse/PerezL17,DBLP:conf/cgo/RountevY14,Jeon2012SymDroidSE} propose different callback control-flow models.
%
%
%
Many previous works, like FlowDroid~\cite{DBLP:conf/pldi/ArztRFBBKTOM14} and Hopper~\cite{DBLP:conf/oopsla/BlackshearCS15}, directly implement the lifecycle of Android components.
While the main intention of these tools is to implement the lifecycle automata, in practice, they also encode some of the effects of callins invoked in the app code in an ad-hoc manner.
%
For example, FlowDroid determines if and where a callback (e.g., \code{onClick}) is registered using a pre-defined list of callin methods and an analysis of the app call graph.
%
Hopper implements the lifecycle callback control flow directly in a static analysis algorithm that efficiently explores the interleaving of Android components.
%
%
In contrast, our work starts from the observation that reasoning about protocol violations requires capturing, in a first-class manner, the effects that invoking a callin has on the future execution of callbacks (and vice-versa).
%

Callback control-flow graphs~\cite{DBLP:conf/icse/YangYWWR15} are graphs of callbacks generated from an application and a manually written model of the framework.
Perez and Le~\cite{DBLP:conf/icse/PerezL17} generate callback control-flow graphs with constraints relating program variables to callback invocations analyzing the Android framework.
Such models can indirectly capture callin effects via the predicates on the program state.
With lifestate, we carefully focus on what is observable at the app-framework interface so that lifestate specifications are agnostic to the internal implementation details of the framework.
%
DroidStar~\cite{DBLP:conf/icse/RadhakrishnaLMM18} automatically learns a callback typestate automaton for an Android object
from
a developer-specified set of transition labels using both callbacks and callins symbols.
Such automata specifically represent the protocol for a single object and, differently from lifestate, their labels are not parameterized messages. A callback typestate is thus a coarser abstraction than lifestate since it cannot express the relationships between different message occurrences that are required to describe multi-object protocols.

%
There exist other classes of framework models that represent different and complementary aspects of
the framework than the callback control flow captured by lifestate.
For example, Fuchs et al.~\cite{fuchs:cs-tr-4991} and Bastani et al.~\cite{DBLP:conf/popl/BastaniAA15}
represent the ``heap properties'' implicitly imposed by the framework.
%
\edgeminer{}~\cite{DBLP:conf/ndss/CaoFBEKVC15} and
\scandal{}~\cite{Kim2017SCANDAL} model the registration of callbacks.
Droidel~\cite{DBLP:conf/pldi/BlackshearGC15} also captures callback registration by modeling the reflection calls inside the Android framework code.
%
Similarly, \pasket~\cite{DBLP:conf/icse/JeonQFFS16} automatically learns
implementations of framework classes
that behave according to particular design patterns.
%
%
%
%

%
While framework models have been extensively used to support static and dynamic analysis, not much attention has been paid to validating that the models soundly capture the semantics of the real framework. Wang et al.~\cite{DBLP:conf/pldi/WangZR16} recognized the problem of model unsoundness---measuring unsoundnesses in three different Android framework models.
Unsoundnesses were found even using a much weaker notion of model validation than we do in this work.
A significant advantage of lifestates is that we can validate their correctness with respect to any execution trace, obtained from arbitrary apps, because they speak generically about the app-framework interface.

There exist several programming languages for asynchronous
event-driven systems, such as Tasks~\cite{DBLP:conf/pepm/FischerMM07}
and P \cite{DBLP:conf/pldi/DesaiGJQRZ13}.
In principle, such languages are general enough to develop event-driven
systems such as Android. The purpose of our formalization \semname{} is instead
to provide a formalization that
captures the app-framework interface.
The protocol verification problem for event-driven
applications is related to typestate verification~\cite{Naeem:2008:TAM:1449955.1449792,joshi+2008:predictive-typestate,fink+2008:effective-typestate},
but it is more complex since it requires reasoning about
the asynchronous interaction of both callbacks and callins.
Dynamic protocol verification is similar in spirit to dynamic event-race detection~\cite{DBLP:conf/pldi/MaiyaKM14,DBLP:conf/pldi/HsiaoPYPNCKF14,DBLP:conf/oopsla/BielikRV15,DBLP:conf/tacas/MaiyaGKM16}, which predicts if there is an event data-race
from execution traces.
However, a lifestate violation differs from, and is not directly comparable to, an event data-race. A lifestate violation could manifest as a data race on a framework-internal field, but more commonly it results from encountering an undesirable run-time state within the framework.

\section{Conclusion}
\label{sec:conclusion}

We considered the problem of specifying event-driven application-programming protocols.
The key insight behind our approach is a careful distillation of what is observable at the interface between the framework and the app.
This distillation leads to the abstract notions of permitted messages from the framework to the app (e.g., enabled callbacks) and prohibited messages into the framework from the app (e.g., disallowed callins). Lifestate specification then offers the ability to describe the event-driven application-programming protocol in terms of this interface---capturing both what the app can expect of the framework and what the app must respect when calling into the framework.
We evaluated our approach by implementing a dynamic lifestate verifier called \toolname{} and showed that the richness of lifestates are indeed necessary to verify real-world Android apps as conforming to actual Android protocols.

\emph{Acknowledgements.}
Many thanks to Edmund S.L. Lam, Chance Roberts, and Chou Yi for help in gathering traces, as well as Alberto Griggio for a convenient tool for running tests. We also thank Aleksandar Chakarov, Maxwell Russek, the Fixr Team, and the University of Colorado Programming Languages and Verification (CUPLV) Group for insightful discussions, as well as the anonymous reviewers for their helpful comments.
This material is based on research sponsored by DARPA under agreement number FA8750-14-2-0263.

\bibliographystyle{plainurl}
\bibliography{conference.short,bec.short,main.short}

\begin{thebibliography}{10}

\bibitem{android-activity-lifecycle}
{Android Developers}.
\newblock The {Activity} lifecycle.
\newblock
  \url{https://developer.android.com/guide/components/activities/activity-lifecycle.html},
  2018.

\bibitem{exercise-monkey}
{Android Developers}.
\newblock {UI/Application} exerciser monkey.
\newblock \url{https://developer.android.com/studio/test/monkey.html}, 2018.

\bibitem{android-topekabug}
{Android Topeka}.
\newblock Crash if rotate device right after press floating action button \#4
  {Topeka} for {Android}.
\newblock \url{https://github.com/googlesamples/android-topeka/issues/4}, 2015.

\bibitem{DBLP:conf/apsec/ArnatovichNTS16}
Yauhen~Leanidavich Arnatovich, Minh~Ngoc Ngo, Hee Beng~Kuan Tan, and Charlie
  Soh.
\newblock Achieving high code coverage in android {UI} testing via automated
  widget exercising.
\newblock In {\em Asia-Pacific Software Engineering Conference (APSEC)}, 2016.

\bibitem{DBLP:conf/pldi/ArztRFBBKTOM14}
Steven Arzt, Siegfried Rasthofer, Christian Fritz, Eric Bodden, Alexandre
  Bartel, Jacques Klein, Yves~Le Traon, Damien Octeau, and Patrick McDaniel.
\newblock {FlowDroid}: Precise context, flow, field, object-sensitive and
  lifecycle-aware taint analysis for {Android} apps.
\newblock In {\em Programming Language Design and Implementation (PLDI)}, 2014.

\bibitem{DBLP:conf/popl/BastaniAA15}
Osbert Bastani, Saswat Anand, and Alex Aiken.
\newblock Specification inference using context-free language reachability.
\newblock In {\em Principles of Programming Languages (POPL)}, 2015.

\bibitem{DBLP:conf/oopsla/BielikRV15}
Pavol Bielik, Veselin Raychev, and Martin~T. Vechev.
\newblock Scalable race detection for {Android} applications.
\newblock In {\em Object-Oriented Programming Systems, Languages, and
  Applications (OOPSLA)}, 2015.

\bibitem{DBLP:conf/oopsla/BlackshearCS15}
Sam Blackshear, Bor{-}Yuh~Evan Chang, and Manu Sridharan.
\newblock Selective control-flow abstraction via jumping.
\newblock In {\em Object-Oriented Programming Systems, Languages, and
  Applications (OOPSLA)}, 2015.

\bibitem{DBLP:conf/pldi/BlackshearGC15}
Sam Blackshear, Alexandra Gendreau, and Bor{-}Yuh~Evan Chang.
\newblock {Droidel}: A general approach to {Android} framework modeling.
\newblock In {\em State of the Art in Program Analysis (SOAP)}, 2015.

\bibitem{boyle-dismissProgress}
Chris Boyle.
\newblock Simon {Tatham's} puzzles.
\newblock
  \url{https://github.com/chrisboyle/sgtpuzzles/blob/658f00f19172bdbceb5329bc77376b40fe550fcb/app/src/main/java/name/boyle/chris/sgtpuzzles/GamePlay.java\#L183},
  2014.

\bibitem{DBLP:conf/ndss/CaoFBEKVC15}
Yinzhi Cao, Yanick Fratantonio, Antonio Bianchi, Manuel Egele, Christopher
  Kruegel, Giovanni Vigna, and Yan Chen.
\newblock {EdgeMiner}: Automatically detecting implicit control flow
  transitions through the {Android} framework.
\newblock In {\em Network and Distributed System Security (NDSS)}, 2015.

\bibitem{DBLP:conf/cav/CavadaCDGMMMRT14}
Roberto Cavada, Alessandro Cimatti, Michele Dorigatti, Alberto Griggio,
  Alessandro Mariotti, Andrea Micheli, Sergio Mover, Marco Roveri, and Stefano
  Tonetta.
\newblock The {nuXmv} symbolic model checker.
\newblock In {\em Computer-Aided Verification (CAV)}, 2014.

\bibitem{fdroid-swiftnotes}
Adrian Chifor.
\newblock Swiftnotes.
\newblock \url{https://f-droid.org/en/packages/com.moonpi.swiftnotes/}, 2015.

\bibitem{kistenstapeln}
{D120}.
\newblock Kistenstapeln.
\newblock \url{https://github.com/d120/Kistenstapeln-Android}, 2015.

\bibitem{DBLP:conf/pldi/DesaiGJQRZ13}
Ankush Desai, Vivek Gupta, Ethan~K. Jackson, Shaz Qadeer, Sriram~K. Rajamani,
  and Damien Zufferey.
\newblock {P:} safe asynchronous event-driven programming.
\newblock In {\em Programming Language Design and Implementation (PLDI)}, 2013.

\bibitem{antennapodbug}
Martin Fietz.
\newblock {FeedRemover}: already running - issue \#1304 -
  {AntennaPod/AntennaPod}.
\newblock \url{https://github.com/AntennaPod/AntennaPod/issues/1304}, 2015.

\bibitem{fink+2008:effective-typestate}
Stephen~J. Fink, Eran Yahav, Nurit Dor, G.~Ramalingam, and Emmanuel Geay.
\newblock Effective typestate verification in the presence of aliasing.
\newblock {\em ACM Trans. Softw. Eng. Methodol.}, 17(2), 2008.

\bibitem{DBLP:conf/pepm/FischerMM07}
Jeffrey Fischer, Rupak Majumdar, and Todd~D. Millstein.
\newblock Tasks: language support for event-driven programming.
\newblock In {\em Partial Evaluation and Program Manipulation (PEPM)}, 2007.

\bibitem{fuchs:cs-tr-4991}
Adam~P. Fuchs, Avik Chaudhuri, and Jeffrey~S. Foster.
\newblock {SCanDroid}: Automated security certification of {Android}
  applications.
\newblock Technical Report CS-TR-4991, University of Maryland, College Park,
  2009.

\bibitem{yamba-git}
Marko Gargenta.
\newblock Yamba.
\newblock
  \url{https://github.com/learning-android/Yamba/blob/429e37365f35ac4e5419884ef88b6fa378c023f8/src/com/marakana/android/yamba/StatusFragment.java},
  2014.

\bibitem{Gargenta:2014aa}
Marko Gargenta and Masumi Nakamura.
\newblock {\em Learning Android}.
\newblock O'Reilly Media, 2014.

\bibitem{osmtracker}
Nicolas Guillaumin.
\newblock {OSMTracker} for {Android}.
\newblock
  \url{https://github.com/nguillaumin/osmtracker-android/blob/d80dea16e456defe5ab62ed8b5bc35ede363415e/app/src/main/java/me/guillaumin/android/osmtracker/gpx/ExportTrackTask.java},
  2015.

\bibitem{DBLP:conf/pldi/HsiaoPYPNCKF14}
Chun{-}Hung Hsiao, Cristiano Pereira, Jie Yu, Gilles Pokam, Satish
  Narayanasamy, Peter~M. Chen, Ziyun Kong, and Jason Flinn.
\newblock Race detection for event-driven mobile applications.
\newblock In {\em Programming Language Design and Implementation (PLDI)}, 2014.

\bibitem{Jeon2012SymDroidSE}
Jinseong Jeon, Kristopher~K. Micinski, and Jeffrey~S. Foster.
\newblock {SymDroid}: Symbolic execution for {Dalvik} bytecode.
\newblock Technical report, Department of Computer Science, University of
  Maryland, College Park, 2012.

\bibitem{DBLP:conf/icse/JeonQFFS16}
Jinseong Jeon, Xiaokang Qiu, Jonathan Fetter{-}Degges, Jeffrey~S. Foster, and
  Armando Solar{-}Lezama.
\newblock Synthesizing framework models for symbolic execution.
\newblock In {\em International Conference on Software Engineering (ICSE)},
  2016.

\bibitem{joshi+2008:predictive-typestate}
Pallavi Joshi and Koushik Sen.
\newblock Predictive typestate checking of multithreaded {Java} programs.
\newblock In {\em Automated Software Engineering (ASE)}, 2008.

\bibitem{facebookbug}
Vladislav Kaplun.
\newblock Update {RequestAsyncTask.java} by kaplad - pull request \#315 -
  facebook/facebook-android-sdk.
\newblock \url{https://github.com/facebook/facebook-android-sdk/pull/315},
  2014.

\bibitem{DBLP:conf/msr/KechagiaS14}
Maria Kechagia and Diomidis Spinellis.
\newblock Undocumented and unchecked: exceptions that spell trouble.
\newblock In {\em Mining Software Repositories, (MSR)}, 2014.

\bibitem{Kim2017SCANDAL}
Jinyung Kim, Yongho Yoon, Kwangkeun Yi, and Junbum Shin.
\newblock {SCANDAL}: Static analyzer for detecting privacy leaks in {Android}
  applications.
\newblock {\em IEEE Mobile Security Technologies (MoST).}, 2017.

\bibitem{DBLP:journals/lisp/Levy06}
Paul~Blain Levy.
\newblock Call-by-push-value: Decomposing call-by-value and call-by-name.
\newblock {\em Higher-Order and Symbolic Computation}, 19(4), 2006.

\bibitem{DBLP:conf/tacas/MaiyaGKM16}
Pallavi Maiya, Rahul Gupta, Aditya Kanade, and Rupak Majumdar.
\newblock Partial order reduction for event-driven multi-threaded programs.
\newblock In {\em Tools and Algorithms for the Construction and Analysis of
  Systems (TACAS)}, 2016.

\bibitem{DBLP:conf/pldi/MaiyaKM14}
Pallavi Maiya, Aditya Kanade, and Rupak Majumdar.
\newblock Race detection for {Android} applications.
\newblock In {\em Programming Language Design and Implementation (PLDI)}, 2014.

\bibitem{verivitatr}
Shawn Meier, Sergio Mover, and Bor{-}Yuh~Evan Chang.
\newblock Lifestate: Event-driven protocols and callback control flow (extended
  version).
\newblock {\em CoRR}, abs/, 2019.

\bibitem{Naeem:2008:TAM:1449955.1449792}
Nomair~A. Naeem and Ondrej Lhot{\'{a}}k.
\newblock Typestate-like analysis of multiple interacting objects.
\newblock In {\em Object-Oriented Programming Systems, Languages, and
  Applications (OOPSLA)}. {ACM}, 2008.

\bibitem{nextgislogger}
{NextGis}.
\newblock {NextGisLogger}.
\newblock \url{https://github.com/nextgis/nextgislogger}, 2017.

\bibitem{onebusawaybug}
OneBusAway.
\newblock {IllegalStateException}: {Fragment} {BaseMapFragment} not attached to
  {Activity} \#570 {OneBusAway}.
\newblock \url{https://github.com/OneBusAway/onebusaway-android/issues/570},
  2016.

\bibitem{DBLP:conf/icse/PerezL17}
Danilo~Dominguez Perez and Wei Le.
\newblock Predicate callback summaries.
\newblock In {\em International Conference on Software Engineering (ICSE)},
  2017.

\bibitem{PingPlusPlus}
PingPlusPlus.
\newblock {Ping Plus Plus}.
\newblock \url{https://github.com/PingPlusPlus/pingpp-android}, 2017.

\bibitem{xxv-androidlifecycle}
Steve Pomeroy.
\newblock The complete {Android} {Activity}/{Fragment} lifecycle v0.9.0.
\newblock \url{https://github.com/xxv/android-lifecycle}, 2014.

\bibitem{DBLP:conf/icse/RadhakrishnaLMM18}
Arjun Radhakrishna, Nicholas~V. Lewchenko, Shawn Meier, Sergio Mover,
  Krishna~Chaitanya Sripada, Damien Zufferey, Bor{-}Yuh~Evan Chang, and Pavol
  Cern{\'{y}}.
\newblock {DroidStar}: callback typestates for {Android} classes.
\newblock In {\em International Conference on Software Engineering (ICSE)},
  2018.

\bibitem{redreaderbug}
{Red Reader}.
\newblock Crash during commenting \#467 {RedReader}.
\newblock \url{https://github.com/QuantumBadger/RedReader/issues/467}, 2017.

\bibitem{gator-toolkit}
A.~Rountev, D.~Yan, S.~Yang, H.~Wu, Y.~Wang, and H.~Zhang.
\newblock {GATOR}: Program analysis toolkit for {Android}.
\newblock \url{http://web.cse.ohio-state.edu/presto/software/}, 2017.

\bibitem{DBLP:conf/cgo/RountevY14}
Atanas Rountev and Dacong Yan.
\newblock Static reference analysis for {GUI} objects in {Android} software.
\newblock In {\em Code Generation and Optimization (CGO)}, 2014.

\bibitem{noveldroid-bug}
sh1ro.
\newblock {NovelDroid}.
\newblock
  \url{https://github.com/sh1r0/NovelDroid/blob/f3245055d7a8bcc69a9bca278fbe890081dac58a/app/src/main/java/com/sh1r0/noveldroid/SettingsFragment.java},
  2016.

\bibitem{DBLP:conf/vstte/SmithC15}
Eric Smith and Alessandro Coglio.
\newblock {Android} platform modeling and {Android} app verification in the
  {ACL2} theorem prover.
\newblock In {\em Verified Software: Theories, Tools, and Experiments (VSTTE)},
  2015.

\bibitem{stackoverflow_fragment_setArguments}
{StackOverflow Post}.
\newblock Got exception: fragment already active.
\newblock
  \url{https://stackoverflow.com/questions/10364478/got-exception-fragment-already-active},
  2012.

\bibitem{stackoverflow_dialog_dismiss}
{StackOverflow Post}.
\newblock Alertdialog creating exception in android.
\newblock
  \url{https://stackoverflow.com/questions/15104677/alertdialog-creating-exception-in-android},
  2013.

\bibitem{click-pause-2}
{StackOverflow Post}.
\newblock {OnClickListener} fired after {onPause}?
\newblock
  \url{https://stackoverflow.com/questions/31432014/onclicklistener-fired-after-onpause},
  2015.

\bibitem{click-after-pause}
{StackOverflow Post}.
\newblock {Android}: click event after {Activity.onPause()}.
\newblock
  \url{https://stackoverflow.com/questions/38368391/android-click-event-after-activity-onpause},
  2016.

\bibitem{strom+1986:typestate:-programming}
Robert~E. Strom and Shaula Yemini.
\newblock Typestate: A programming language concept for enhancing software
  reliability.
\newblock {\em IEEE Trans. Software Eng.}, 12(1), 1986.

\bibitem{audiobug}
Matthias Urhahn.
\newblock {AudioBug}.
\newblock \url{https://github.com/d4rken/audiobug}, 2017.

\bibitem{DBLP:conf/pldi/WangZR16}
Yan Wang, Hailong Zhang, and Atanas Rountev.
\newblock On the unsoundness of static analysis for {Android} {GUIs}.
\newblock In {\em State of the Art in Program Analysis (SOAP)}, 2016.

\bibitem{DBLP:conf/icse/YangYWWR15}
Shengqian Yang, Dacong Yan, Haowei Wu, Yan Wang, and Atanas Rountev.
\newblock Static control-flow analysis of user-driven callbacks in {Android}
  applications.
\newblock In {\em International Conference on Software Engineering (ICSE)},
  2015.

\end{thebibliography}

\end{document}